\documentclass[usenatbib]{mnras}

\usepackage{footnote}
\usepackage{amsmath}
\usepackage{amssymb}
\usepackage{xcolor}
\usepackage{graphicx}
\usepackage{pdflscape}
\usepackage{url}
\usepackage{txfonts}
\usepackage[normalem]{ulem}
\usepackage{bm}		
\usepackage[OT1]{fontenc}
\usepackage{booktabs}

\newcommand{\gtsim}{\mbox{{\raisebox{-0.4ex}{$\stackrel{>}{{\scriptstyle\sim}}$}
}}}
\newcommand{\ltsim}{\mbox{{\raisebox{-0.4ex}{$\stackrel{<}{{\scriptstyle\sim}}$}
}}}
\newcommand{\cib}{\ensuremath{\,{\rm Jy}\,{\rm deg}^{-2}}}

\DeclareGraphicsExtensions{.pdf,.png,.jpg}

\title[ALMACAL IX]{ALMACAL IX: multi-band ALMA survey for dusty star-forming galaxies and the resolved fractions of the cosmic infrared background}

\author[J.~Chen et al.]{Jianhang~Chen,$^{\! 1}$\thanks{Email: Jianhang.Chen@eso.org}
R.\,J.~Ivison,$^{\! 1}$\thanks{Email: Rob.Ivison@eso.org}
Martin~A.~Zwaan,$^{\! 1}$\thanks{Email: mzwaan@eso.org}
Ian~Smail,$^{\! 2}$
Anne~Klitsch,$^{\! 3}$
C\'eline~P\'eroux,$^{\! 1,4}$
\newauthor
Gerg\"{o}~Popping,$^{\! 1}$
Andrew~D.~Biggs,$^{\! 5}$
Roland~Szakacs,$^{\! 1}$
Aleksandra~Hamanowicz$^{6}$
and Claudia~Lagos$^{7,8,9}$\\
$^{1}$European Southern Observatory (ESO), Karl-Schwarzschild-Strasse~2, D-85748 Garching, Germany\\
$^{2}$Centre for Extragalactic Astronomy, Department of Physics, Durham University, South Road, Durham DH1 3LE\\
$^{3}$DARK, Niels Bohr Institute, University of Copenhagen, Jagtvej 128, 2200 Copenhagen, Denmark\\
$^{4}$Aix Marseille Université, CNRS, LAM (Laboratoire d’Astrophysique de Marseille) UMR 7326, 13388, Marseille, France\\
$^{5}$UK Astronomy Technology Centre, Royal Observatory, Blackford Hill, Edinburgh EH9 3HJ\\
$^{6}$Space Telescope Science Institute, 3700 San Martin Dr., Baltimore MD 21218, USA\\
$^{7}$International Centre for Radio Astronomy Research (ICRAR), M468, University of Western Australia, 35 Stirling Hwy, Crawley, WA 6009, Australia\\
$^{8}$ARC Centre of Excellence for All Sky Astrophysics in 3 Dimensions (ASTRO 3D), Australia\\
$^{9}$Cosmic Dawn Center (DAWN)
}

\date{Accepted 2022 October 14. Received 2022 October 7; in original form 2022 August 5}
\pubyear{2022}                                  
 
\begin{document}
\label{firstpage}
\maketitle

\begin{abstract}
Wide, deep, blind continuum surveys at submillimetre/millimetre (submm/mm) wavelengths are
required to provide a full inventory of the dusty, distant Universe.  
However, conducting such surveys to the necessary depth,
with sub-arcsec angular resolution, is prohibitively
time-consuming, even for the most advanced submm/mm telescopes.
Here, we report the most recent results from the ALMACAL project,
which exploits the `free' calibration data from the Atacama Large
Millimetre/submillimetre Array (ALMA) to map the lines of sight 
towards and beyond the ALMA calibrators. 
ALMACAL has now covered 1,001 calibrators, with a total sky coverage around 0.3 deg$^2$, 
distributed across the sky accessible from the Atacama desert, 
and has accumulated more than 1,000\,h of integration.
The depth reached by combining multiple visits to each field 
makes ALMACAL capable of searching for faint,
dusty, star-forming galaxies (DSFGs), with detections at multiple frequencies to constrain the emission mechanism.
Based on the most up-to-date ALMACAL database, we report the detection of 186 DSFGs with flux densities down to $S_{870\mu\text{m}}\!\sim$~0.2~mJy,
comparable with existing ALMA large surveys but less susceptible to cosmic variance.
We report the number counts at five wavelengths between 870\,$\mu$m and 3\,mm, 
in ALMA bands 3, 4, 5, 6 and 7, providing a benchmark for 
models of galaxy formation and evolution.
By integrating the observed number counts and the best-fitting functions, 
we also present the resolved fraction of the cosmic infrared background (CIB)
and the CIB spectral shape.
Combining existing surveys, ALMA has currently resolved about half of the CIB in the submm/mm regime.
\end{abstract}


%

\section{Introduction}

The cosmic infrared background (CIB) -- covering mid-infrared to
submillimetre/millimetre (submm/mm) wavelengths -- comprises an important component of the energy emitted by galaxies over the history of the Universe \citep{Hauser1998, Fixsen1998}.
Along with the optical background, it represents the energy produced by the formation
and evolution of galaxies, and all related processes, across all cosmic time.
Since the first measurements of the CIB, a primary goal in astrophysics
has been to identify the sources responsible for it, and thus to 
understand the lifetime energy budget of the Universe \citep[e.g.][]{Cooray2002,Lagache2005, Dole2006,PlanckCollaboration2014}.

Several decades after the discovery of the CIB, 
various galaxy populations are known to make significant contributions.
Among them, strongly star-forming galaxies with total infrared luminosities larger than $10^{11}$\,L$_{\odot}$ \citep{Sanders1996} are thought to be the main contributors, with galaxies at different redshifts contribute to different parts of the CIB.
In the mid- and far-infrared (-IR), the CIB largely comprises luminous and ultraluminous IR galaxies (LIRGs and ULIRGs) at $z<1.5$ \citep{Elbaz2002}.
In the submm/mm wavelength regime, the CIB is dominated by dusty, star-forming galaxies (DSFGs) at higher redshifts ($z\gtrsim 1.5$).
In the submm/mm bands, 
dimming due to increasing redshift is compensated by stronger dust emission
as the observing frequency traces rest frequencies progressively closer to the peak of a typical dust spectral energy distribution (SED). 
This makes the submm/mm wavelength regime particularly well suited to the detection of dust-rich galaxies at high redshift \citep[e.g.][]{Blain1993, Smail2002}

The first submm-selected galaxies (SMGs, the brightest DSFGs) were found by continuum surveys at 850\,$\mu$m with the 15-m James Clerk Maxwell Telescope (e.g.\ \citealt*{Smail1997};
\citealt{Barger1998,Hughes1998,Eales1999}),  which ushered in the era of submm cosmology.
Since then, SMGs have been studied extensively at different wavelengths, using radio, X-ray and optical telescopes to trace their properties \citep[e.g.][respectively]{Ivison2002,Alexander2005,Chapman2005}.
New instruments and associated surveys have increased the family of SMGs selected at the original wavelength \cite[e.g.][]{Chen2016,Geach2017}, 
at longer wavelengths \citep[$\lambda>1$\,mm, e.g.][]{Scott2012, Magnelli2019}, as well as at shorter wavelengths, e.g.\ 
with the {\it Herschel Space Observatory} \citep[$\lambda<500\,\mu$m, e.g.][]{Eales2010, Oliver2010}.

A significant advance came with the advent of the Atacama Large Mm/submm Array (ALMA), 
which was immediately able to pinpoint and even spatially resolve many of the earlier bright, single-dish-selected SMGs \citep[e.g.][]{Hodge2013,Ikarashi2015,Simpson2015a,Stach2019}.
Some especially bright lensed sources proved to be ideal targets for detailed studies of their internal structures and physical states \citep[e.g.][]{Geach2018,Rizzo2020,Dye2022},
while other bright SMGs were resolved into multiple sources, leading to the discovery of proto-clusters of extreme DSFGs
\citep[e.g.][]{Ivison2013,Oteo2018,Miller2018}.

ALMA also offers the sensitivity and spatial resolution necessary to push below the confusion limit imposed by single-dish imaging, allowing the discovery of much fainter DSFGs.
As happened with the single-dish telescopes that came before, two methods have been used to undertake surveys for DSFGs: 
long integrations in conventional `deep fields' \citep[e.g.][]{Walter2016,Dunlop2017,Umehata2018} 
and the targeting of lensing clusters to probe smaller areas where the gravitational magnification is high \citep[e.g.][]{Gonzalez-Lopez2017,Laporte2021}, enabling the discovery of faint DSFGs in the epoch of re-ionisation \citep[e.g.][]{Laporte2017,Tamura2019,Fudamoto2021a}.

To create a full inventory of dusty galaxies, we need deep, wide and blind surveys.
Before ALMA, single-dish telescopes had already covered 10s to 100s of square degrees of the sky and found many thousands of bright FIR/submm/mm sources.
However, the diffraction limit prevented us from delving below the intensely star-forming ULIRG regime, aside from a small number of strongly lensed systems.
It was also necessary to make model-based corrections to the counts to deal with issues associated with source blending, and a significant portion of bright SMGs found by single-dish telescopes were later resolved into multiple sources by interferometers \citep[e.g.][]{Ivison2007,Wang2011,Karim2013}.
Using ALMA, it has been possible to extend the submm/mm detection limit down to the sub-mJy level. 
Even with the small number of antennas available in Cycle~0,
ALMA pushed down to $S_{870\mu\text{m}}\!\sim$~0.4\,mJy in the 
Extended {\it Chandra} Deep Field South \citep{Karim2013,Hodge2013}.
Since then, extensive follow-up campaigns have been carried out for the bright SMGs found in earlier single-dish surveys \citep[e.g.][]{Weiss2013,Miettinen2017,Brisbin2017,Cowie2018,Stach2019,Simpson2020}.
Meanwhile, blind ALMA surveys have been continuously enlarging the survey area and/or improving the detection limit \citep[e.g.][]{Hatsukade2016,Hatsukade2018,Walter2016,Dunlop2017,Umehata2018}.
Very recently, the ALMA Spectroscopic Survey in the Hubble Ultra Deep Field (ASPECS) Large Programme \citep{Walter2016} has achieved a sensitivity
of 10\,$\mu$Jy\,beam$^{-1}$ \citep{Gonzalez-Lopez2019,Gonzalez-Lopez2020} and 
although ALMA is not optimised for large sky coverage,
the GOODS-ALMA project has observed the 72\,arcmin$^2$ Great Observatories Origins Deep Survey South field (GOODS-South)
in two different array configurations \citep{Franco2018, Gomez-Guijarro2022}.
Considerable time has also been invested in the well-known legacy fields, which have also inspired dedicated data mining projects \citep[e.g.][]{Ono2014,Zavala2018,Liu2019}.

All these ALMA surveys have helped to constrain the number counts and the properties of DSFGs.
Number counts -- the projected galaxy surface density as a function of flux density -- represent the most basic measurement we can glean from such observations.
They therefore provide a simple test of the validaty of models of galaxy formation and evolution.
Indeed, both semi-analytic models \citep[e.g.][]{Lacey2016,Somerville2012,Lagos2020} and post-processing models in hydrodynamic simulations \citep[e.g.][]{Shimizu2012,McAlpine2019,Cowley2019,Lovell2021}
have struggled to reproduce the number counts of DSFGs.
In addition, accurate and unbiased number counts are essential to resolve the CIB
and determine the contributions of different galaxy populations to the total CIB.
However, due the very high demand for ALMA time
it has become difficult to go much deeper or wider, 
and it has proved even more difficult to justify covering the same areas at multiple submm/mm wavelengths.

\begin{figure*}
   \centering
   \resizebox{0.9\hsize}{!}{\includegraphics{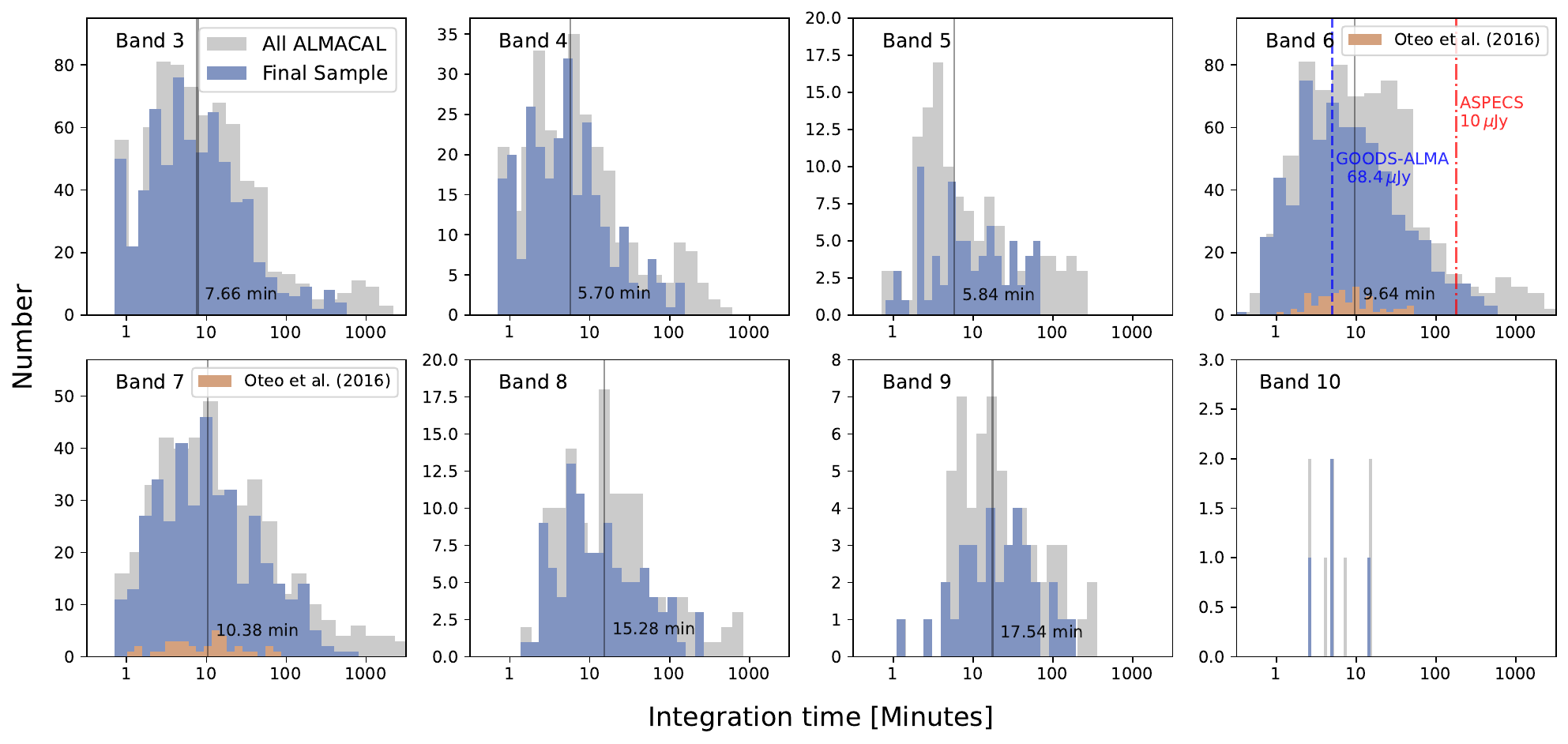}}
   \caption{Statistics of on-source time of ALMACAL observations from ALMA bands 3 to 10. For every panel, the abscissa is the accumulated total on-source time in units of minutes. The ordinate is the number of fields in every time step. We show the total data available in each band and the data used in this work. Some of them have been dropped because of calibration errors or strong residuals after calibrator removal (see \S\ref{sec:data_selection}). In bands 6 and 7 we also show the statistics for the data used by \citet{Oteo2016}, illustrating the immense increase since that time. Both the number of fields and the sensitivity in other bands have increased, making it possible to expand the number counts to a wider flux density range and to multiple ALMA bands. In band 6, the two vertical lines show the sensitivity reached by the ASPECS \citep{Gonzalez-Lopez2020} and GOODS-ALMA surveys \citep{Franco2018,Gomez-Guijarro2022}. ALMACAL provides a good compromise between their sensitivity and sky coverage.}
   \label{fig:statistics_comparison}
\end{figure*}

ALMACAL is a novel submm/mm survey that exploits the `free'
data that ALMA must collect to ensure its observations can be processed to make spectra, images and cubes with accurate positions, polarisations and flux densities.
With ALMACAL, we survey the immediate vicinity of each calibrator, which are typically blazars at $z<1$ \citep{Bonato2018,Klitsch2019}.
Since $\approx20$ per cent of all ALMA observing time is spent on calibration, 
ALMACAL is already competitive with the widest and deepest submm/mm surveys. 
It does come with an obvious disadvantage: 
the lack of ancillary data at depths comparable with 
classic deep fields, such as the Cosmic Evolution Survey (COSMOS) 
or GOODS. 
However, it also comes with several major advantages. 
Number counts of ALMACAL are largely immune to cosmic variance, since it covers a great many pointings scattered across the observable sky,
and the presence of an in-beam calibrator allows for perfectly calibrated submm/mm/radio 
imaging with a very high dynamic range.
The blind detections share the same spatial and spectral set-ups
as the science targets, whatever they may have been, which makes it possible 
to search for spectral lines (in absorption or emission) and to study morphologies at the very highest spatial resolution, up to 20\,mas \citep[see][]{Oteo2017}.
Meanwhile, these fields will be visited repeatedly by future submm/mm and radio interferometers, 
which will keep improving the sensitivities in these fields at different wavelengths. 

In our earliest attempt at mining the ALMA calibration data,
\citet{Oteo2016} described the first search for DSFGs in ALMACAL.
With multi-band data available for most of the fields, they derived dual-band number counts, in bands 6 (1.2\,mm) and 7 (870\,$\mu$m), using the data collected before 2015.
Since then, both the number of calibrators and their on-source 
integration times have grown by more than an order of magnitude. 
Using the new data, \citet{Klitsch2020} reported the first number counts in ALMA band 8, at 650\,$\mu$m.
Besides number counts, searches for molecular absorption or emission lines have put constrains on the evolution of molecular gas density over cosmic time (\citealt{Klitsch2019}; Hamanowicz et al.~2022, submitted).
In this work, we extend the number counts of \citet{Oteo2016} 
and \citet{Klitsch2020} to the most up-to-date ALMACAL dataset, covering ALMA bands 3, 4, 5, 6 and 7, at wavelengths from 3\,mm down to 870\,$\mu$m, respectively.

The paper is structured as follows: in \S\ref{sec:almacal}
we introduce the ALMACAL project and its results to date;
\S\ref{sec:observation_analysis} details our data analysis, 
including source detection, corrections for flux boosting, sample completeness, 
effective area, source classification and selection bias.
\S\ref{sec:results} presents the final source catalogue and the number counts in five different ALMA bands.
\S\ref{sec:discussion} contains relevant discussion and comparison between our number counts with literature and model predictions. 
The various fractions of the CIB resolved by ALMA are also presented in \S\ref{sec:discussion}.
We summarise our key results in \S\ref{sec:conclusions}.
Throughout this paper, we follow the terminology used in the review by \citet{Hodge2020}. 
We use SMGs to denote the classical submm-selected galaxies with $S_{870\mu\text{m}}\!\ge$~1.0\,mJy and use DSFGs to include all the dusty, star-forming galaxies we can detect via ALMA continuum observations.
In this paper, we assume a $\Lambda$CDM cosmology with $H_0$=67.7 and $\Omega_m=0.31$ \citep{PlanckCollaboration2020}.
\begin{figure*}
   \centering
   \resizebox{0.9\hsize}{!}{\includegraphics{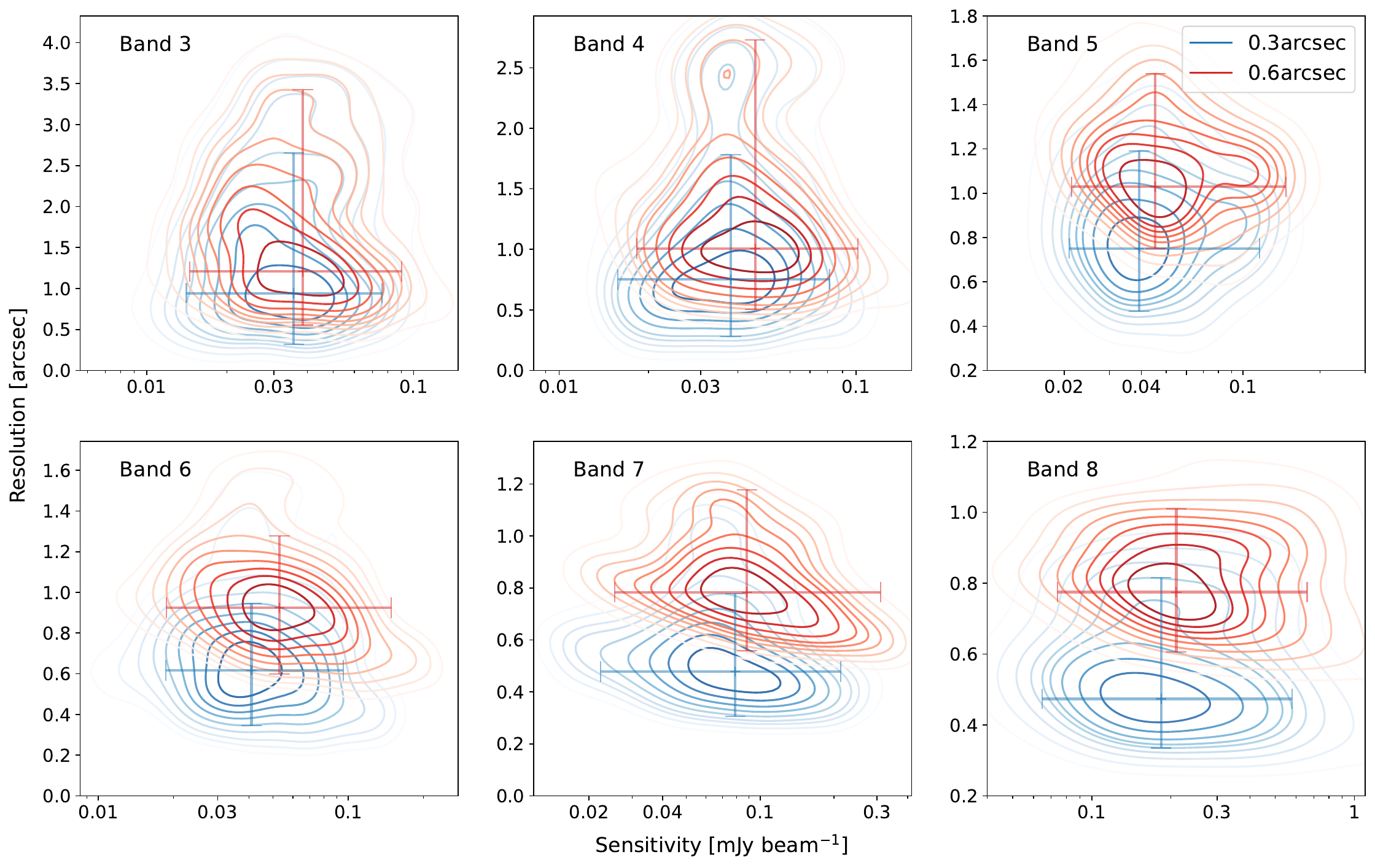}}
   \caption{Density distribution of sensitivity versus resolution for all the combined images in each ALMA band. We show the distributions for each set of tapered images in different colours. The lowest contours include 90 per cent of the data points and decrease in steps of 10 per cent. Our two $uv$ tapers ensure the images have at least 0.3$''$ and 0.6$''$ spatial resolution, respectively. In the final images, most of the sources should remain unresolved.}
   \label{fig:sensitivity_vs_beamsize}
\end{figure*}

\section{ALMACAL}\label{sec:almacal}

ALMACAL\footnote{https://almacal.wordpress.com} aims to exploit all ALMA calibrator scans for science \citep{Zwaan2022}.
Running now for more than ten years, ALMACAL has already accumulated more than 1,000\,h of data.
As of March 2020, ALMACAL includes 1,001 calibrators.
About 97 per cent of them are classified as blazars \citep{Bonato2018}, which are quasars -- active galactic nuclei -- whose jets are oriented very close to our line of sight, such that relativistic beaming makes them extremely bright. 
The calibrators are spread all over the sky and used to calibrate science targets local to them.
Depending on their brightness, compactness and flux stability, they can be used to calibrate bandpass, gain (complex amplitude and phase), flux density and polarisation.
In a typical ALMA scheduling block (SB), two or more calibrators will be observed, along with the science targets, so calibrators share the same instrumental configuration as the science targets.
The rich data buried in the calibration observations make them far more useful than their original intention.

One intriguing example is the calibrator named J1058+0133, one of the brightest blazars close to Cosmic Evolution Survey (COSMOS) field.
\citet{Oteo2017} found two $z=3.4$ SMGs behind this calibrator.
ALMA has invested hundreds of hours in the COSMOS field, which makes J1058+0133 one of the most frequently visited calibrators, observed in total for around 150\,h.  
With the rich range of configurations that ALMA has employed in the COSMOS field, \citet{Oteo2017} were able to create a high signal-to-noise multi-band image of the two SMGs, self-calibrated on the timescale of 1\,sec using the in-beam calibrator, J1058+0133, with 20\,mas spatial resolution.
Additionally, multiple CO lines have been identified in this system, yielding a remarkable spectral line energy distribution, making this amongst the most extraordinary datasets gathered for any SMG.

A detailed description of the ALMACAL data retrieval and calibration pipeline was given in the first ALMACAL paper, \citet{Oteo2016}, which we summarise briefly here.
All the data reduction is carried out using the Common Astronomy Software Application \citep[{\sc casa} --][]{McMullin2007a}.
Firstly, the calibration data are requested from the ALMA science archive.
Next, we apply the standard calibration pipeline, following the \emph{ScriptForPI.py} Python script for each project and splitting out the calibrator data.
For calibrators that lack flux density calibration -- mostly the bandpass calibrators -- their flux densities are recovered from the internal flux tables in the data delivery package.
After that, two cycles of self-calibration are performed in the multi-frequency synthesis (mfs) image. 
The first cycle is focused on phase-only solutions; the second cycle corrects the amplitude and phase together.
Next, a point-source model of the calibrator is removed from the $uv$ data by the {\sc casa} internal tool {\sc uvmodelfit} to make science-ready data.
Finally, the fully calibrated measurements are re-binned to a channel width of 15.6\,MHz.
After these pre-processing steps, the scans contain the fully calibrated data, with the central calibrator subtracted out.

This specially designed pipeline takes full advantage of the bright, central source to do self-calibration, avoids possible effects due to variability, and it is straightforward to combine calibrated data from many different projects.
The dynamic range of images made for calibrators exceeds $10^4$, where the background sources or calibrator jets are typically 100 times weaker than the calibrators themselves.
After removing the central, bright calibrator from each dataset, we can also easily identify the observations that are suitable for combination.
At the moment, we simply reject any problematic data (see \S\ref{sec:data_selection} for details). 
Due to the immense size of the dataset, it is not realistic to re-calibrate all the problematic observations manually.

ALMA has invested significant observing time in various deep cosmology fields \citep[see a summary in][]{Hodge2020}.
ASPECS is currently the deepest blind survey, covering 2.9\,arcmin$^{2}$ with an r.m.s.\ sensitivity that reaches $\sigma_{\rm 1.2mm}\approx9.3\,\mu$Jy\,beam$^{-1}$ \citep{Gonzalez-Lopez2020}.
Meanwhile, GOODS-ALMA represents the largest blind survey, with sky coverage of 72.4\,arcmin$^{2}$, with a much shallower sensitivity, $\sigma_{\rm 1.2mm}\approx70\,\mu$Jy\,beam$^{-1}$ \citep{Gomez-Guijarro2022}.
ALMA follow-up of bright sources from the single-dish surveys is the most efficient way to probe the very brightest SMGs and such an approach has been employed extensively in the existing legacy fields \citep[e.g.][]{Weiss2013,Stach2019,Simpson2020}.
ALMACAL manages to combine the best depth and sky coverage -- see \S\ref{sec:effective_area}.
In Fig.~\ref{fig:statistics_comparison}, we show the on-source time for all our calibrators in the different ALMA receiver bands.
Compared with \citet{Oteo2016} in bands 6 and 7, the total on-source time and the number of available calibrators have both increased dramatically.
Ultimately, ALMACAL will go deeper and wider than any ALMA blind survey, unless a significant strategic investment is made.

\section{Observations and Analysis}\label{sec:observation_analysis}

We detail the steps we have taken regarding data selection and imaging in \S\ref{sec:data_selection}, then present the methods and steps taken to construct the source catalogue in \S\ref{sec:source_detecting} and \ref{sec:flux_density}.
After that, we derive the correction functions for our survey, including those for flux deboosting (\S\ref{sec:flux_boosting}), completeness (\S\ref{sec:completeness}), effective area (\S\ref{sec:effective_area}) and effective wavelength (\S\ref{sec:effective_wavelength}).
Finally, we classify our sources in \S\ref{sec:source_classification} and discuss selection biases in \S\ref{sec:selection_bias}.

\begin{figure*}
   \centering
   \resizebox{0.9\hsize}{!}{\includegraphics{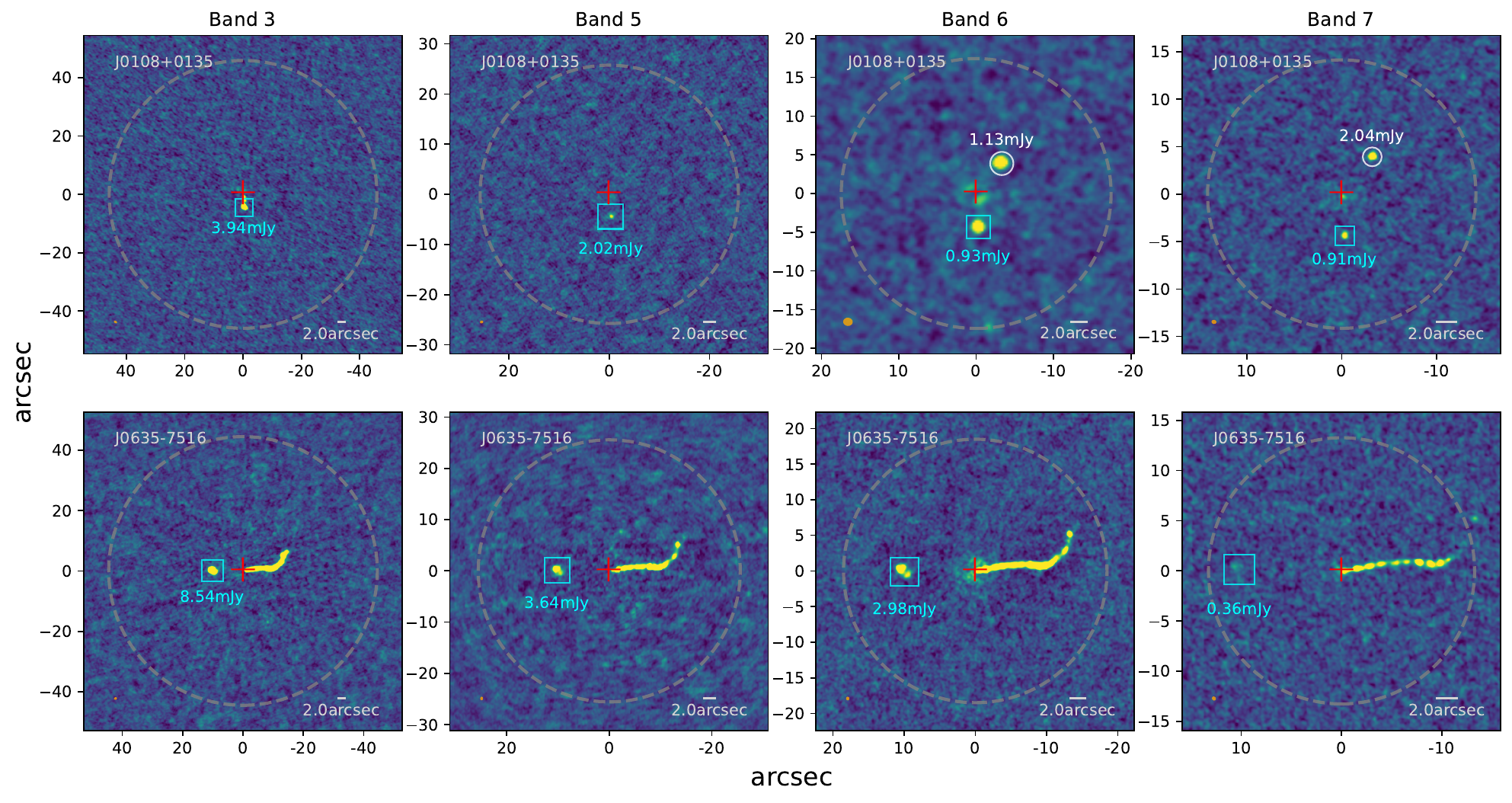}}
   \caption{Two examples of detections in ALMACAL in different bands. The wavelength is decreasing from left to right. In every image, the central bright calibrator has been removed -- its position is marked with a red cross -- and all the available observations have been combined. The dashed grey circle is the FoV adopted in our work, which is 1.8\,$\times$\,FWHM of the respective primary beam. In the top row, the images in ALMA bands 4, 5, 6 and 7 towards the field J0108+0135 are shown, where one SMG (marked with a white circle) and one jet (marked with a cyan square) have been found. The SMG has higher flux densities at higher frequencies, while, in contrast, the flux density of the radio jet decreases. In the bottom row, the images from the field J0635$-$7516 are shown, where two-sided extended radio jets have been discovered. We have classified each detection based on its spectral index and morphology.}
   \label{fig:example_targets}
\end{figure*}

\subsection{Data selection and imaging}\label{sec:data_selection}

A rose has its thorns. 
The bright calibrator at the centre of the field can be helpful for self-calibration but at times it brings problems.
Firstly, the calibrator is not always a point source; blazars have jets, and they are not always (or have not always been) oriented along the line of sight; they can be slightly resolved, or amplitude and phase errors can remain after the calibration steps.
All these issues can introduce residuals after point-source subtraction of the central calibrator.
To avoid residuals corrupting the final, combined image, we applied two imaging cycles.

In the first imaging cycle, we image the visibilities of every single dataset after point-source removal, then we visually inspect all the images, discarding the ones with incorrect calibration or strong residuals.
This step alone led to the loss of around half of our data, but it ensured the best quality of the final combined images.
Having removed the poor quality data, we re-scale the weighting of different observations with the {\sc casa} task, {\sc statwt}, then combine the observations of each field using the task, {\sc concat}.
All the images were made using {\sc tclean} in {\sc casa} version 5.7.0 in `mfs' mode, which combines all the frequency channels into a single continuum image.

The second imaging cycle use the combined, re-scaled visibilities to create science-ready images.
One of the major goals of ALMACAL is to search for DSFGs in the calibrator fields.
From previous ALMA follow-up of known, bright DSFGs, we expect that they have compact dust morphologies, with a median size $\loa 0.5$\,arcsec \citep[FWHM,][]{Ikarashi2015, Simpson2015, Gullberg2019}.
To optimise the sensitivity to compact sources, we adopted a natural weighting scheme and the Hogbom \citep{Hogbom1974} deconvolution algorithm to clean our images.
In addition, to avoid resolving DSFGs we further tapered the visibilities with Gaussian kernels of 0.3 and 0.6\,arcsec.
We show the final resolution versus sensitity in Fig.~\ref{fig:sensitivity_vs_beamsize}, our uvtapers ensure the images have at at least 0.3$''$ and 0.6$''$ spatial resolution, respectively.
These two tapered images were used for source detection, and to quantify the fraction of missing flux density.
During the deconvolution, we first estimated the sensitivity of the final combined visibilities using the function {\sc sensitivity} in 
{\sc casa} \emph{Analysis Utilities}\footnote{https://casaguides.nrao.edu/index.php/Analysis\_Utilities} 
and used this as the threshold in {\sc tclean}.
In every major cleaning cycle, the built-in {\sc auto-multithresh} algorithm in {\sc tclean} was used to search for secure emission and determine the clean regions in the residual maps.
We cleaned the image within an area defined as $1.8\times$ the FWHM of the primary beam. 
After cleaning, {\sc impbcor} was used to correct the primary beam response.
In the following steps, as a convenience, the image {\it without} the primary beam correction was used for source detection, while the primary-beam-corrected image was used for flux density measurements.

\begin{figure*}
   \centering
   \resizebox{0.9\hsize}{!}{\includegraphics{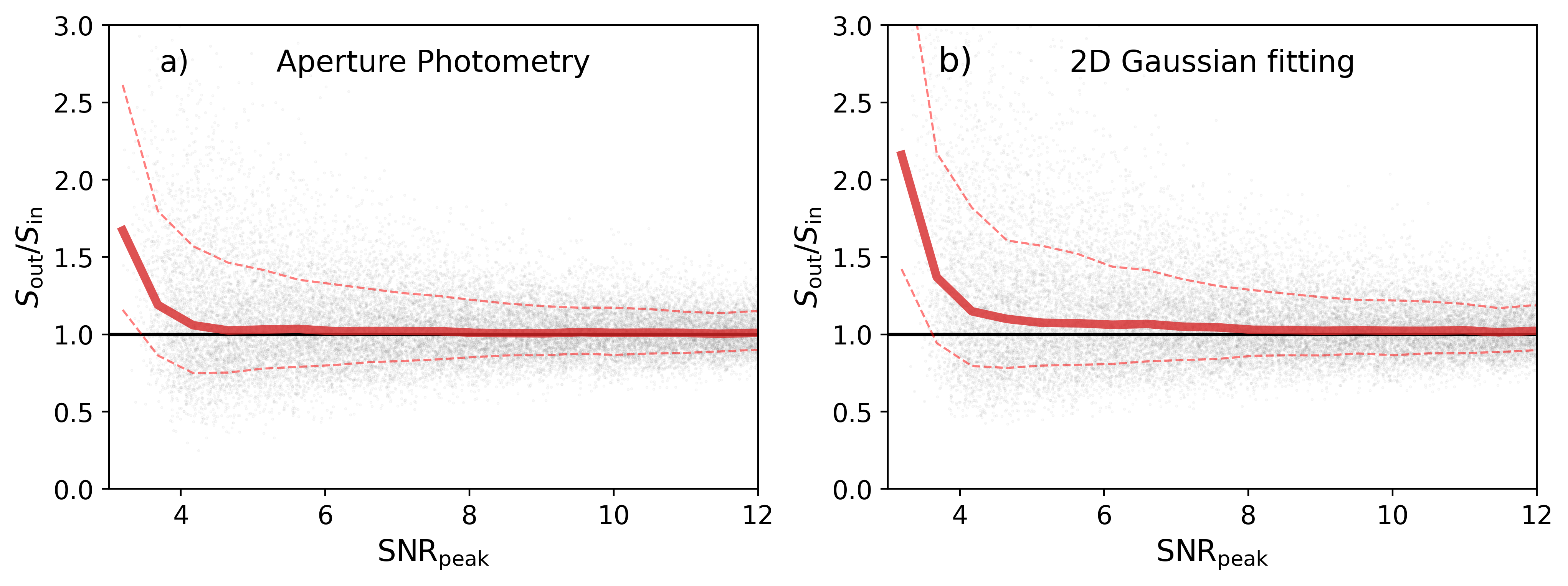}}
   \caption{Flux boosting as a function of peak SNR. The boosting effects have been estimated from the recovered flux density of artificial sources (0.2\,arcsec FWHM) randomly injected in the observed image. The abscissa is the peak SNR of the injected source, and the ordinate is the ratio between the measured flux density, $S_{\rm out}$, and the injected flux density, $S_{\rm in}$. The two methods used for flux measurements are shown in the two panels. Panel a) is used aperture photometry; panel b) is used 2-dimensional Gaussian fitting. The red solid line is the median value in different SNR bins, the red dashed lines enclose 68 per cent of the points in each bin. Our simulations indicate that different flux measuring methods will lead to slightly different deboosting functions, where the Gaussian fitting tends to capture more positive noise during the fitting. In our final catalogue, we report the two measurements corrected by their own de-boosting functions.}
   \label{fig:flux_boosting}
\end{figure*}

\subsection{Source detection}\label{sec:source_detecting}

Following our two earlier number counts papers, \citet{Oteo2016} and \citet{Klitsch2020}, we performed the source detection using {\sc sextractor} as it can deal easily with complex source structures \citep{Bertin1996}.
The search for detections is made in the images before primary beam correction, corresponding to a signal-to-noise image.
We detected residual signals from the calibrator in some fields, so we masked the central part of each map to a radius of 2\,arcsec.
This radius is $2\times$ larger than the masking used in band 8 \citep{Klitsch2020} but the loss is small for the larger fields of view (FoV) at the longer wavelengths explored here.
To improve the reliability of the source detection in ALMACAL, we applied the source-finding algorithm in the two tapered images simultaneously.
We search sources with SNRs higher than $3\sigma$ and then only accept detections with peak SNR higher than 5$\sigma$ in at least one of the two tapered images.
\footnote{Most of our detections ($>$90 per cent) have been found in both of the two tapered images. Those detections found in just one of the tapered images, with a counterpart in another ALMA band, are also classified as robust detections; those without counterparts in any other ALMA band are marked as uncertain. The variation of the final number counts that include these uncertain sources are discussed in \ref{appendixfig:number_counts_perturbation}}
Our adoption of a 5$\sigma$ final detection cut prevents us from finding very faint sources, but ensure our detected sources are robust.
In Fig.~\ref{fig:example_targets}, we show the source detections at different wavelengths in two example fields.

We searched for sources within $1.8\times$ the FWHM of the primary beam, which corresponds to 15 per cent of the peak sensitivity.
This represents a reasonable trade-off, yielding a large total effective area whilst maintaining reasonable sensitivity.
The final positions of all the detections are defined by their signal-to-noise centroid \citep{Bertin1996}.

\subsection{Flux density}\label{sec:flux_density}

We measured the flux densities of our detections using aperture photometry.
Aperture photometry was undertaken using the {\sc Python} package, {\sc photutils}, with elliptical apertures. 
The size of the aperture is important: selecting too large or too small an aperture introduces uncertainties into the total flux density.
Since most of our sources are unresolved, we use the synthesised beam as the shape of our aperture.
Based on our testing, an aperture size that is twice the synthesised beam with an appropriate aperture correction delivers the most reliable flux density.
The aperture correction assumes an unresolved point source.
To better capture all the flux for our detections, we use the 0.6$''$ tapered images as the primary images to measure their flux densities.
The flux difference between the two $uv$ tapered images are small, less than 10 per cent, supporting our assumption that most of emissions from our detections remain unresolved.

We also report the flux densities from 2-dimensional Gaussian fitting.
The Gaussian fitting was achieved by {\sc astropy.modeling} with the {\sc Gaussian2D} model and the {\it LevMarLSQFitter} fitting algorithm.
Before applying the fitting algorithm, a cut-out is taken from the original image, measuring $5\times$ the major axis of the synthesised beam.
During fitting, the amplitude, axial ratio, positional angle could change freely.
The centre of the Gaussian function is allowed to shift by $\pm$0.1\,arcsec relative to the photometric centre.
The flux density was calculated by integrating the best-fitting Gaussian.

The sensitivity of every ALMACAL field is not uniform, decreasing outwards from the phase centre.
This means sources need to be brighter at larger radii to be detectable.
Because the primary-beam correction scales up the flux density and noise by the same factor, the peak SNRs of detections remain the same.
In the following two sections, we discuss flux boosting and sampling completeness as a function of peak SNR, which is convenient to apply to all our fields.

\subsection{Flux deboosting}\label{sec:flux_boosting}

The well-known effect that we must consider when measuring accurate flux densities is the so-called `flux boosting' of faint sources \citep{Hogg1998a}.
It has two origins \citep[e.g.][]{Coppin2006,Casey2014}.
First, we are more likely to detect faint sources that have been scattered towards higher flux densities by random noise (Eddington flux boosting), an effect that can be mitigated to some extent by choosing a high SNR detection threshold.
Second, where there is a rapid increase in the number of sources as we delve fainter, i.e.\ when the source counts are steep, then the resolution element of a telescope may include additional sources that are individually fainter than the detection threshold. 
The latter issue has been common for single-dish submm/mm telescopes, because of their typically large diffraction limit, $\gtsim 10$\,arcsec, and because the source counts of DSFGs at $S_{\rm 850\mu m}\gtsim$6\,mJy are steep \citep{Simpson2020}.

However, the blending effects should be very much less significant for deep, high-resolution interferometric observations, such as those obtained using ALMA.
The spatial resolution of ALMA is considerably higher than that of single-dish telescopes, which makes the confusion noise significantly lower.
At the same time, the number counts of DSFGs have been found to flatten at fainter flux densities \citep[e.g.][]{Stach2018,Gonzalez-Lopez2020}.
The combination of these two differences with historic work ensures that blending effects are smaller for interferometer-based work.
Nevertheless, we have tested the degree of flux boosting affecting our detections via bespoke Monte Carlo simulations in all our fields.

To preserve the noise characteristics of our fields, as well as any undetected faint sources, we injected artificial sources directly into our observed images.
Before we begin, we clean the image and run our source-finding algorithm to search for detections, masking any before passing the image to the simulation.
For each cycle of the simulation, we generated 20 artificial sources for each field.
The artificial sources were randomly assigned flux densities such that their SNRs lay between 2--20\,$\sigma$.
Their positions were also randomly assigned, within the $1.8\times$ FWHM area of the primary beam.
Next, we removed all close pairs (mutual distance $<3$arcsec) of artificial sources to avoid unnatural source blending.
After this pruning, the artificial sources were convolved with the synthesised beam and added to the image.
This process was repeated $1,000\times$ for each field, to gather a statistically significant sample of simulated sources.
We then repeat all the steps for different source sizes.
We model the shape of the injected sources with 2-D Gaussians, with the intrinsic size varying across a grid of five different sizes (point sources with FWHM = 0.1, 0.2, 0.3, 0.6\,arcsec).
Finally, we searched these images for detections in the same way as we do for our sources (see \S\ref{sec:flux_density} and \ref{sec:source_detecting}), using the same methods to measure their flux densities.

We also tested the injection of artificial sources into the $uv$ visibilities.
Similar to the image-plane simulation, we also randomly generate 20 sources per cycle per field.
Then, ALMA task {\sc ft} was used to convert the point sources into visibilities and ALMA task {\sc uvsub} is used to add them into the visibilities.
The two simulations give consistent results, but conducting simulations in the visibilities is time-consuming, especially considering that we need simulations for different fields, so we chose to conduct all the simulations in the image plane for all the ALMACAL fields.

Fig.~\ref{fig:flux_boosting} shows the effects of flux boosting as a function of SNR.
We only show the results for our fiducial source size of 0.2\,arcsec.
The flux boosting also varies for different source sizes, but the flux boosting converges to a single boosting function above $\text{SNR}_{\rm peak} \ge 5$ (see  \S\ref{appdixsec:simulation_fluxboosting}).
Moreover, size measurements are problematic at low SNRs, which complicates any size-based correction.
As a compromise, for simplicity, we focus on the flux boosting for sources with a fixed size (FWHM=0.2\,arcsec) in our analysis.

The flux boosting effect is slightly different for different flux-measuring methods.
We found a smaller boosting factor for aperture photometry than with the 2-D Gaussian fitting. 
For aperture photometry, we did not find noticeable boosting for bright sources (SNR\,$>\!5\sigma$);
while for 2-D Gaussian fitting, the boosting effects are still not negligible for sources with SNR up to 10.
This boosting factor is consistent with \citet{Oteo2016}, whose results were based on two-dimensional Gaussian fitting with CASA task, {\sc imfit}.
The key difference between aperture photometry and Gaussian fitting, as deployed here, is that the latter allows the centroid to move slightly ($\pm 0.1$\,arcsec) and has the freedom to vary the source shape, such that it will suffer more from Eddington flux boosting.
However, Gaussian fitting is more robust than aperture photometry in crowded regions.
In our final catalogue, both of the flux densities are reported and corrected by their own deboosting function.
In the following analysis we use the flux density from aperture photometry, but changing to Gaussian fitted flux density does not change our results significantly.

\begin{figure}
   \centering
   \resizebox{\hsize}{!}{\includegraphics{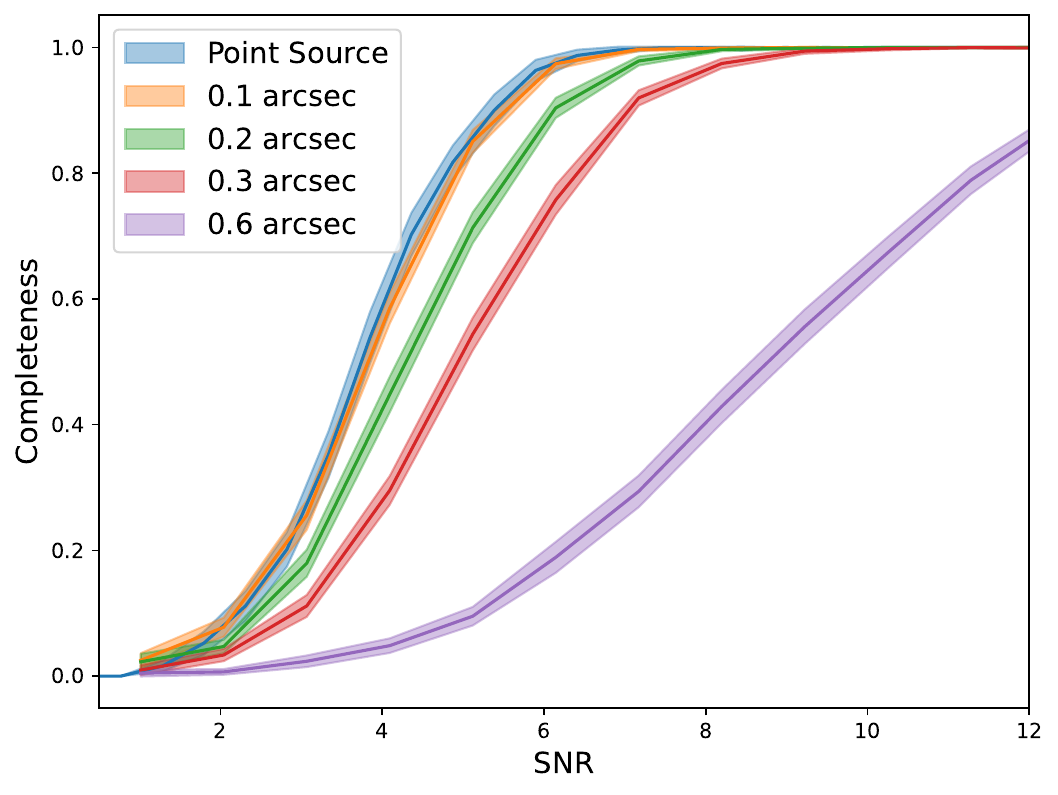}}
   \caption{Sample completeness as a function of peak SNR for various source sizes (in FWHM). The larger the source size, the lower the completeness at a given SNR. However, source size with SNR\,$<$\,10 cannot be  measured robustly. Based on size distribution of DSFGs from current surveys at similar sensitivity, we adopted 0.2$''$ as the fiducial source size to correct the sample completeness.
   We caution that our survey is less sensitive to very extended sources (FWHM\,$>0.6$\,arcsec).}
   \label{fig:completeness}
\end{figure}

\begin{figure}
   \centering
   \resizebox{\hsize}{!}{\includegraphics{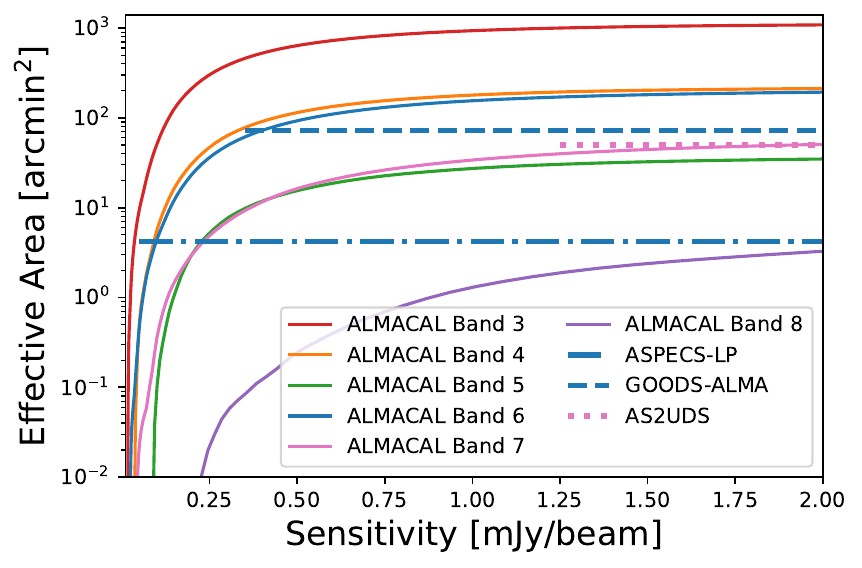}}
   \caption{Effective area as a function of sensitivity ($5\sigma$) for different ALMA bands. The sensitivity reached in each field is different, and decreases with increasing distance to the phase centre within each pointing. These effects lead to the final effective area changing with sensitivity. In addition to ALMACAL, we also show the sensitivities and effective areas of ASPECS \citep{Gonzalez-Lopez2020} and GOODS-ALMA \citep{Gomez-Guijarro2022} conducted in ALMA band 6, and the AS2UDS survey \citep{Stach2019} in ALMA band 7. Compared with existing dedicated ALMA surveys, ALMACAL offers a good balance between sensitivity and effective area and will continue to be an essential complement to existing or ongoing blind surveys.}
   \label{fig:effective_area}
\end{figure}

\begin{figure*}
   \centering
   \resizebox{0.9\hsize}{!}{\includegraphics{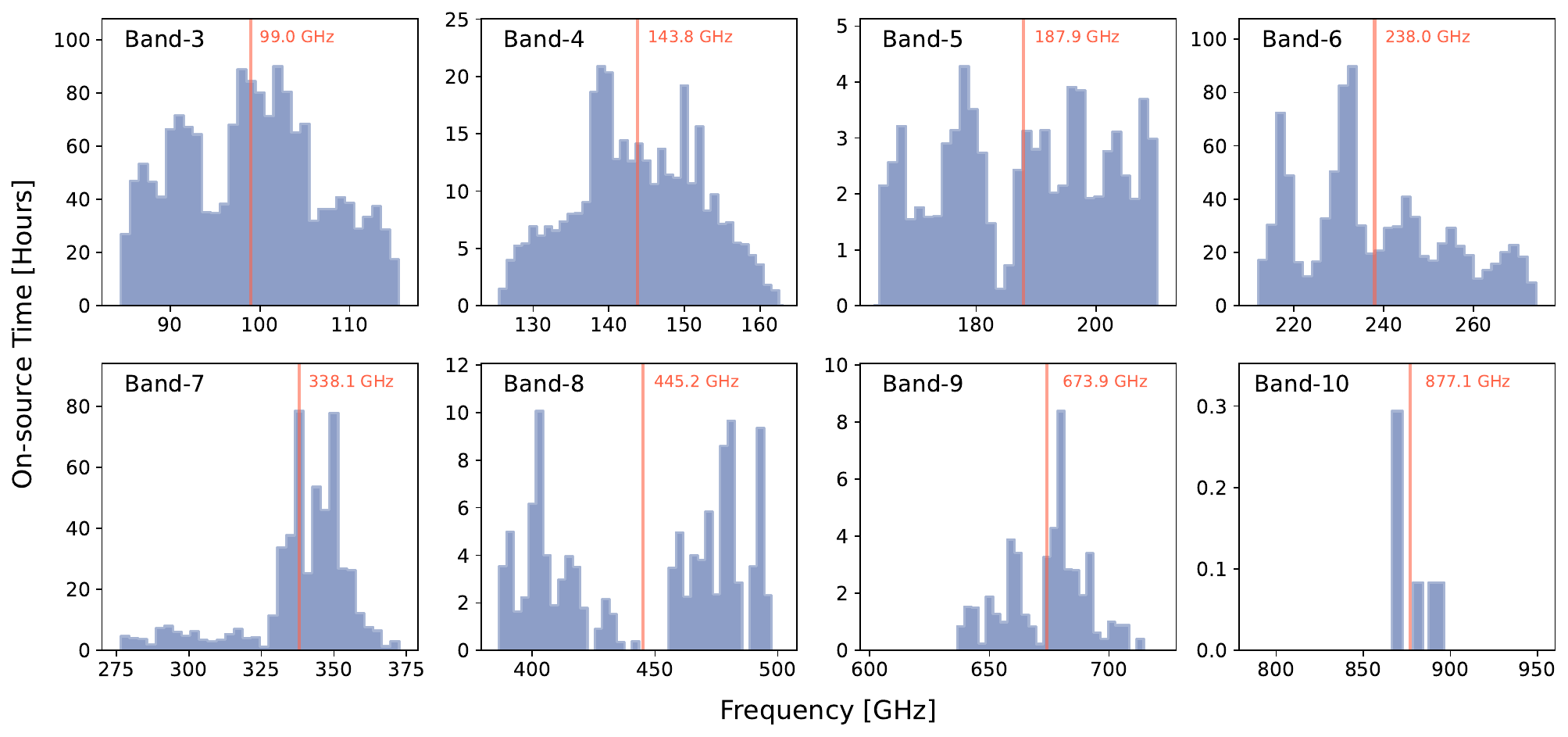}}
   \caption{ALMACAL frequency coverage in the different ALMA bands. The abscissa is the frequency coverage in each ALMA band. The ordinate is the total integration time for the selected observations at each frequency. For each band, the vertical red line is the time-weighted mean frequency. In contrast to existing ALMA blind surveys, the frequency coverage of ALMACAL is much wider. To make meaningful comparisons with literature results, number counts have been re-scaled to the characteristic frequencies based on the SED of a modified blackbody (see \S\ref{sec:number_counts} for more detail).}
   \label{fig:spw_statistics}
\end{figure*}

\subsection{Completeness \& reliability}\label{sec:completeness}

We use the same simulation as discussed in \S\ref{sec:flux_boosting} to derive the completeness correction for our catalogue.
We apply the source-finding procedures to these simulated images, using the methods described in \S\ref{sec:source_detecting}.
An injected source was marked as recovered if it was matched by a detection within the synthesised beam, and was otherwise marked as missed.
If a source was detected without a corresponding injected source, it was marked as a false detection.

As shown in Fig.~\ref{fig:completeness}, the resulting completeness varies for different source sizes.
Size-based corrections have been discussed in the literature \citep[e.g.][]{Bethermin2020}, but the intrinsic source size cannot be well constrained for sources with SNR\,$<10$ \citep{Simpson2015}.
We therefore adopt the completeness correction derived from our fiducial source size (FWHM=0.2\,arcsec).
Besides a robust size measurement, the intrinsic size distribution of DSFGs is needed to fully correct the completeness, which requires future observational effort \citep[e.g.][]{Gullberg2019,Smail2021}.
We did not find many extended sources in the existing surveys with sensitivities and resolutions similar to ALMACAL.
Therefore, we stick to the fiducial completeness correction but caution that this as a limitation of our analysis.

With a 5$\sigma$ detection threshold, we found no spurious detections in our simulations.
The spurious fraction goes down to zero around SNR\,$\sim4.7$ in most of our fields, hence the adopted $5\sigma$ threshold will give us a clean and robust sample.

\subsection{Effective area}\label{sec:effective_area}

Unlike surveys made via mosaics of uniform ALMA pointings, the effective area of ALMACAL changes with sensitivity.
Firstly, different fields have different total integration times, which leads to different sensitivities.
On top of this, the sensitivity decreases with increasing distance from the phase centre, as per the primary beam response.
To quantify this effect, we measured the effective area as a function of peak flux density.
For every field, the effective area for a particular flux density is dictated by the radius at which its SNR drops to 5$\sigma$.
The maximum effective radius is limited by the $1.8\times$ FWHM for source detection.
Both the SNR threshold and the maximum radius are the same as those used for source detection.
We performed the same calculation for all of the available fields to determine the total effective area at different peak flux densities.

Fig.~\ref{fig:effective_area} shows the total effective area as a function of peak flux density for the different ALMA bands.
Due to the larger FoV at longer wavelengths, and its popularity with ALMA users, ALMA band 3 has the largest effective area, close to 700\,arcmin$^{2}$ at a flux density of 0.1\,mJy.
It is followed by bands 4 and 6.
The ALMA FoV in band 6 is much smaller than in band 4, but band 6 has more observations -- 39 per cent of all the ALMACAL observing time. 
Bands 5 and 7 have similar effective areas at flux densities larger than 0.3\,mJy, but band 7 goes much deeper.
Besides ALMACAL, we also plot the sky coverage of other pointed and blank-field surveys  \citep{Gonzalez-Lopez2020,Franco2018,Stach2019,Gomez-Guijarro2022}.
By comparison, ALMACAL offers a good balance between sensitivity and effective area, and will continue to be an essential complement to blind surveys.

\citet{Klitsch2020} estimated the cosmic variance of band 8 footprints from ALMACAL following the methodology described in \citet{Driver2010}.
They found the cosmic variance is less than 5 per cent level in ALMACAL band 8 observations.
All the bands with longer wavelengths have at least $10\times$ larger sky coverage, which should suffer from less than 1 per cent cosmic variance.

\subsection{Effective wavelength}\label{sec:effective_wavelength}

The spectral coverage of ALMACAL is much more complex than that of previous ALMA surveys where the observations were typically carried out at a fixed frequency.
ALMA calibrators share the same configurations as their science targets, 
such that the same calibrator observed by different projects will potentially have different spectral configurations.
Fig.~\ref{fig:spw_statistics} shows the frequency coverage of ALMACAL from ALMA band 3 to 10.
The vertical lines are the exposure-time-averaged mean frequency for each band.
For example, the observations in band 6 are spread across the whole band.
The exposure-weighted mean frequency in band 6 is 237.95\,GHz (1.26\,mm), which is slightly lower than the most commonly used 250\,GHz (1.2\,mm).
By comparison, the coverage in band 7 is simpler, with most of the observations undertaken at around 340\,GHz.  Its mean frequency is 338.08\,GHz (887\,$\mu$m).
The effects of these differences need to be corrected before a comparison is made with literature results, as discussed in \S\ref{sec:number_counts}.

\begin{figure}
   \centering
   \includegraphics[width=\linewidth]{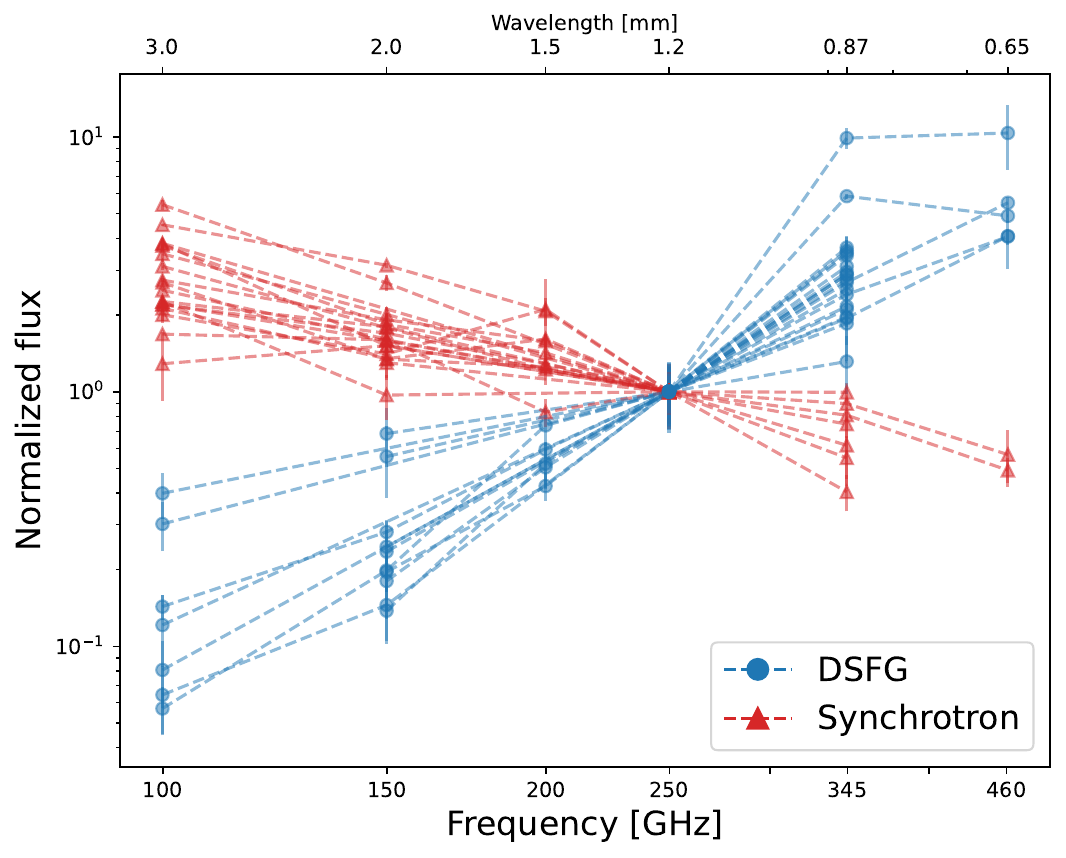}
   \caption{The submm/mm SED of a few confirmed DSFGs and Synchrotron sources. Only the sources detected in band 6 plus at least two more bands are shown here. All the SEDs have been normalized by their band 6 flux densities. We see that DSFGs have positive spectral indices with their fluxes rising at higher frequencies, while the Synchrotron sources typically have negative spectral indices with their fluxes decreasing at higher frequencies.}%
   \label{fig:source_SEDs}
\end{figure}

\begin{figure}
   \centering
   \resizebox{\hsize}{!}{\includegraphics{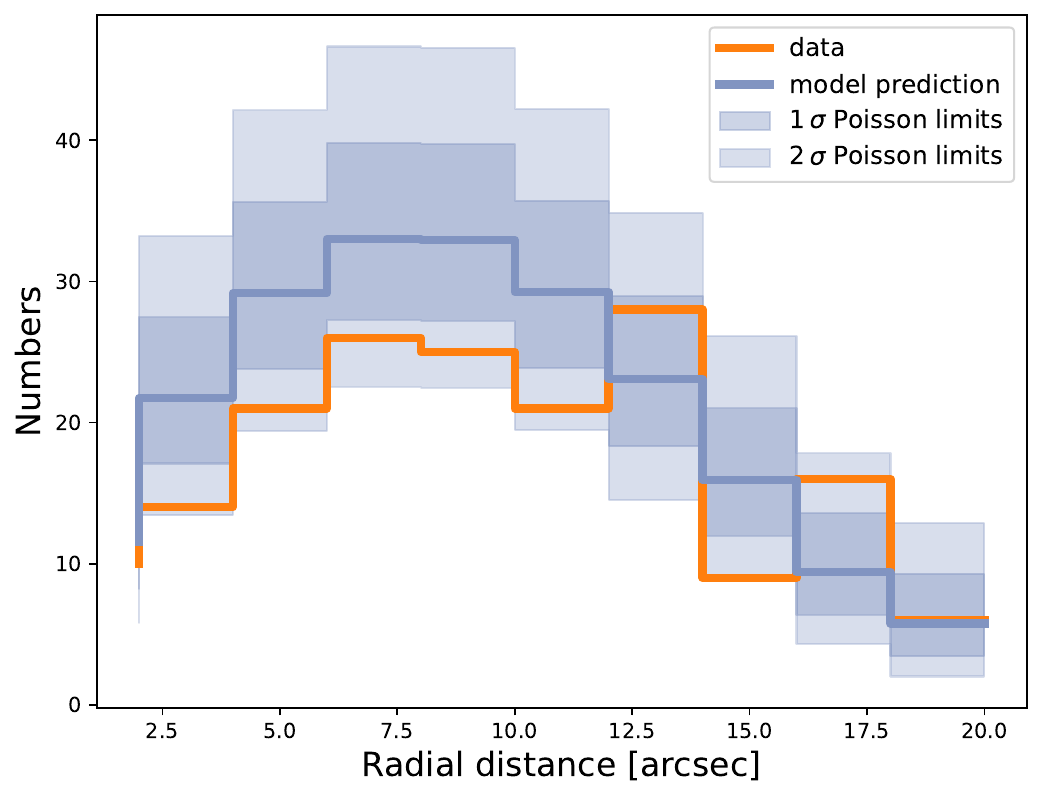}}
   \caption{Radial distribution of DSFGs detected in band 6 in the ALMACAL survey. The abscissa is the radial distance of the detections from their field centre. The orange histogram is the radial distribution of the confirmed DSFGs. The blue histogram is the predicted DSFGs distribution based on the expected number counts from ASPECS \citet{Gonzalez-Lopez2020}. The blue and light blue shadows are the $1\sigma$ and $2\sigma$ Poisson variations of the model predictions. Small differences between ALMACAL and ASPECS are expected. However, the radial distribution of our DSFGs is generally consistent with the expected of randomly distributed DSFGs, which indicates that the thermal sources are not clustered around the blazars.}%
   \label{fig:radial_distribution}
\end{figure}

\begin{table*}
  \caption{Statistical summary of ALMACAL detections.}
  \label{tab:statistics}
  \begin{tabular}{ccccccccc}
    \hline\hline
    ALMA & Ref.~$\lambda$ & Sky coverage & On-source time & $\langle\nu\rangle$ & Number of detections & DSFGs & Synchrotron & Unclassified sources    \\ 
    Band & (mm) & (arcmin$^{2}) $ & (h) & (GHz) &  &   & \\ 
    \hline

    3 & 3.00 & 817 & 250.6  & 99.0 & 63 & 8 & 44 & 12\\
    4 & 2.00 & 157 & 53.1 & 143.75 & 54 & 20 & 23 & 11\\
    5 & 1.50 & 27 & 20.8 & 200 & 21 & 10 & 9 & 2\\
    6 & 1.20 & 149 & 344.7 & 250 & 228 & 132 & 54 & 42\\
    7 & 0.87 & 45 & 275.2 & 345 & 130 & 93 & 22 & 15\\
    8 & 0.65 & 5 & 52.7 & 460 & 17 & 13 & 3 & 1\\
    9 & 0.45 & 1 & 20.9 & 666 & 2 & 2 & 0 & 0\\
    10 & 0.35 & 0.1 & 0.5 & 857 & 0 & 0 & 0 & 0\\
    \hline
    total & -- & 1201.1 & 1018.5 & -- & 371 & 186 & 102 & 83\\
    \hline\hline
    \label{tab:summary}
  \end{tabular}
\end{table*}

\subsection{Source classification}\label{sec:source_classification}

Sources detected by ALMACAL are typically either thermal in nature, e.g.\ continuum emission from dust in star-forming galaxies, or non-thermal in nature, e.g.\ synchrotron emission from radio jets associated with the calibrator or other radio sources in the field.
Thanks to the multi-band coverage of ALMACAL, these emission mechanisms can be separated by their spectral index, $\alpha$, where $S_{\nu}\propto\nu^{\alpha}$.
For a typical DSFG at $z\sim 2$, we are probing the Rayleigh-Jeans tail of emission from warm dust with $\alpha\approx+3.5$ \citep{Ivison2010,Swinbank2010}, so the emission becomes brighter at higher frequencies.
For radio AGN and jets, on the other hand, the flux density typically declines at higher frequencies or stays flat, such that $\alpha \ltsim 0$.
We categorised a source as a DSFG if its multi-band flux densities are consistent with a dust SED, and as synchrotron source if its spectral index instead betrays synchrotron emission \citep[see also][]{Klitsch2020}.

We show the SED from examples of confirmed sources in Fig.~\ref{fig:source_SEDs}.
To obtain the dust SEDs of sources covered by different ALMA bands, we extend the flux measurement to all the available bands whenever a detection is confirmed.
The aperture size and aperture correction are calculated using the same way we do for true detections.
We show all the measurements that have reliable flux densities (SNR $\ge$ 3) in at least three ALMA bands in Fig.~\ref{fig:source_SEDs}. 
In general, DSFGs have distinguishable SEDs from radio sources.
The median $\alpha$ in our sample is $\alpha=3.2\pm0.3$ and $\alpha=-0.9\pm0.1$ for DSFGs and Synchrotron, respectively.

We commonly encounter the radio jets of our blazar calibrators, which are not always perfectly aligned with the line of sight.
Most are relatively compact, beginning close to the calibrator; they tend to be much more prominent in bands 3 and 4 and are easily identified by their positions and elongated morphologies.
However, extended radio jets have been found in about 10 per cent of the fields.
These jets could be resolved into several blobs and morphologically they resemble point sources at higher frequencies.
With detections in multiple bands, these jets can be easily excluded via their spectral indices and their morphological alignment (see the second example in Fig.~\ref{fig:example_targets}).

The classification of an ALMACAL source is uncertain if the source is only covered by one ALMA band.
In such cases, we first we cross-matched with radio images from the Monitoring Of Jets in Active galactic nuclei with VLBA Experiments (MOJAVE)\footnote{https://www.physics.purdue.edu/astro/MOJAVE/}, Faint Images of the Radio Sky at Twenty-cm  \citep[FIRST --][]{Becker1995}, the NRAO VLA Sky Survey \citep[NVSS --][]{Condon1998} and the NRAO VLA Archive Survey
(NVAS)\footnote{https://www.vla.nrao.edu/astro/nvas/}. 
If the source has been detected by the radio surveys given their typical depth, it is classified as synchrotron.
The additional radio images are crucial for sources found in band 3 due to the generally wider FoV and higher sensitivity.
Examples of our synergetic classification, combining data from ALMA and the VLA, are illustrated in Fig.~\ref{appendixfig:VLA_and_ALMA}.
In band 3, we are able to confirm 14 per cent more of the total sources with the help of radio surveys.

For those unclassified sources lacking radio coverage, because of their small numbers ($<20\%$), we did not find a large difference in number counts by including these uncertain sources, so in the main context we focus on reliable SMGs that are confirmed by at least two ALMA bands or one ALMA band with the radio images available.
Details of the flux density distribution, radial distance distribution and the contribution to the number counts of the unclassified sources can be found in Fig.~\ref{appendixfig:B6_histogram} and Fig.~\ref{appendixfig:number_counts_perturbation}.
Source classifications are available from our on-line version of the catalogue.

\begin{figure*}
  \centering
  \resizebox{0.9\hsize}{!}{\includegraphics{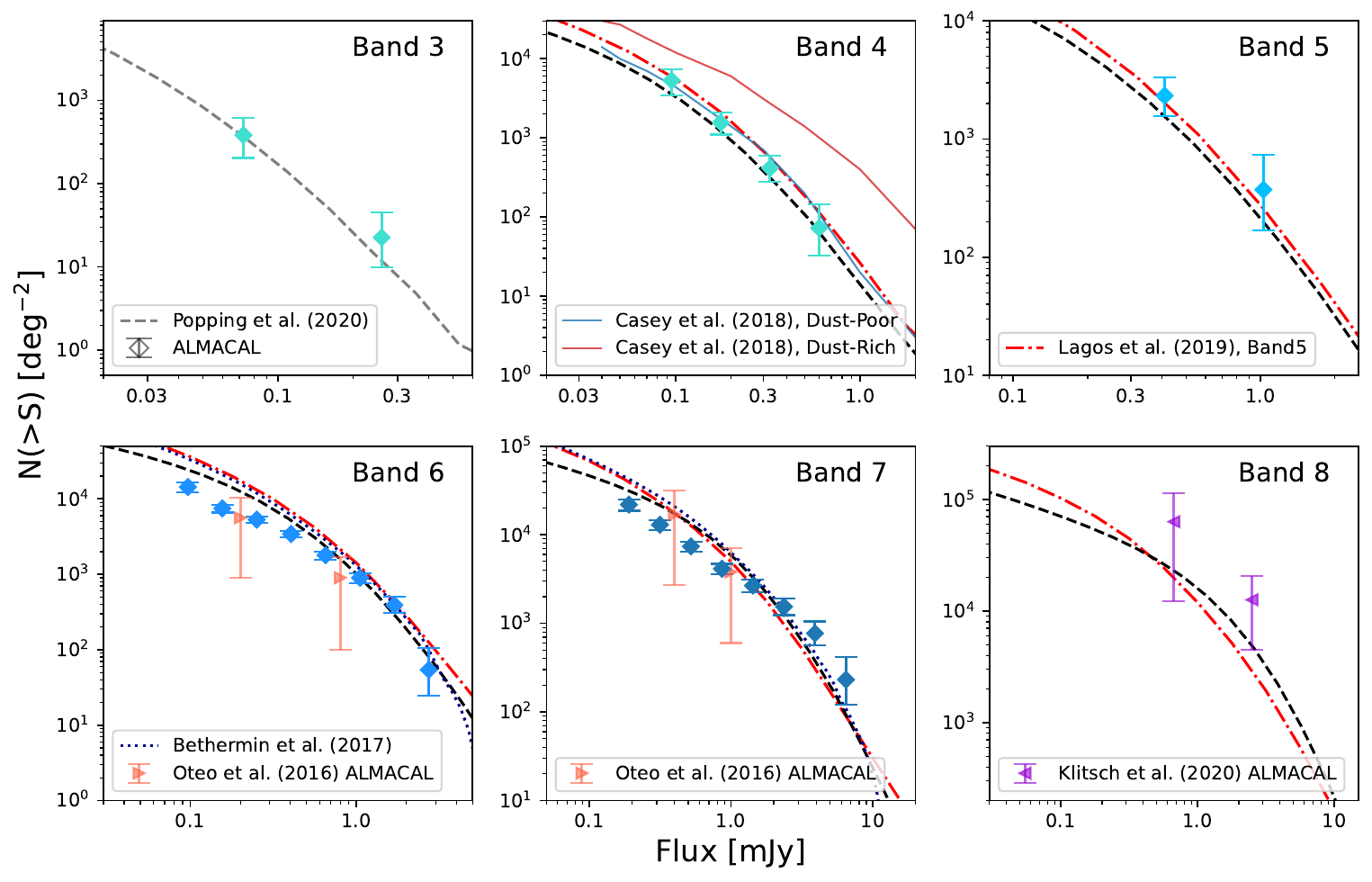}}
  \caption{Multi-band number counts from ALMACAL. The number counts reported in this work are displayed in each panel. In bands 6 and 7, we also show the first ALMACAL number counts by \citet{Oteo2016}; in band 8, we adopt the results from \citet{Klitsch2020}. We also show model predictions of the number counts at various wavelengths, including the semi-analytical model from \citet{Lagos2020} and the semi-imperical models from \citet{Popping2020} and \citet{Bethermin2017}. A successful model should be able to explain the number counts at the different wavelengths simultaneously. The most recent semi-analytical and semi-empirical models are broadly consistent with the number counts of ALMACAL at bright fluxes, but show apparent differences at fainter limits in band 6 and 7 (see also Fig.~\ref{fig:number_counts_B7} and Fig.~\ref{fig:number_counts_B6}), which need to be confirmed by the future surveys. Therefore, multi-wavelength number counts are a powerful tool to validate those models.}
  \label{fig:number_counts_all}
\end{figure*}

\subsection{Selection bias}\label{sec:selection_bias}

ALMACAL is not a truly blind survey. 
The pre-selection of the calibrator fields can bias our survey if the central calibrator has influenced the detection of nearby sources.
Since roughly 97 per cent of the calibrators are blazars, the blazar activity may have been triggered by an interaction, perhaps also giving rise to a nearby starburst.
In addition, the blazar host galaxy may act as a gravitational lens.
However, a blazar is bright mainly due to the beaming effect, which does not always traces the most massive galaxies and dark matter halos. 
The redshift distribution of the ALMACAL blazars peaks at $0.5<z<1.0$ \citep{Bonato2018}, well short of the typical redshifts to DSFGs \citep{Brisbin2017,Aravena2020,Dudzeviciute2020}.
Moreover, we note that a systematic search for galaxy over-densities around three distant quasars with known companions failed to find associated DSFGs \citep{Meyer2022}.

To further quantify the possible biases that might be introduced by the central bright sources, in Fig.~\ref{fig:radial_distribution} we compare the radial distribution of all our DSFGs with the prediction of the ALMA blind survey, ASPECS.
We used the best-fitting number counts from ASPECS as the count model and then created a mock survey in the manner of ALMACAL.
We show the radial distribution from band 6 where the number of detections is the highest.
If the same density of sources detected by ASPECS is recovered here, scattered randomly throughout our fields, we would have unbiased DSFG positions, regardless of the central blazar.
We made mock observations by using the same observational set-ups as ALMACAL, covering the same effective area at different flux densities, and the  uncertainty in the radial distribution is given by the Poisson error.
Fig.\,\ref{fig:radial_distribution} shows the $1\,\sigma$ and $2\,\sigma$ Poisson confidence levels indicating that the radial distribution of our detections is broadly consistent with ASPECS.
In particular, we do not see any excess of detections near the central calibrators, indicating there is little evidence for clustering of the DSFGs around the blazars (see also \S\ref{appdixsec:number_counts_innerouter} and Fig.~\ref{appendixfig:inner_outer} for the difference in number counts between inner and outer parts of all the fields.).

\section{Results}\label{sec:results}

In this section, we present the source catalogue and the number counts at different wavelengths.

\subsection{Source catalogue}

Our classification of sources is carried out by the procedures presented in \S\ref{sec:source_classification} and the numbers of different detections are summarised in Table~\ref{tab:statistics}.
In total, we have detected 371 sources, including 186 secure DSFGs with multi-band confirmations of their purity and spectral indices.
The largest number of detections is in ALMA band 6, where we have detected 228 sources, including 132 DSFGs.
This is followed by band 7, where 93 DSFGs from 130 detections have been confirmed.
The detection rate in band 4 is also promising, with 20 secure DSFGs detected.
Band 5 has ten DSFGs, which enables us to derive the number counts at 1.5\,mm for the first time.
In band 3, sky coverage is significantly larger than higher frequency bands and they are typically deeper, which makes the radio images much more important to classify the detections.
Luckily, around half of the ALMACAL band 3 footprints have been covered by VLA archival surveys.
Combining ALMACAL with the VLA archive, we have confirmed eight DSFGs in band 3.
Band 9 has only two detections found in 1~arcmin$^2$ -- too few to meaningfully constrain the number counts.
For ALMA band 8, we adopt the results of \citet{Klitsch2020}, which already benefited from the updated version of ALMACAL.

As we can see here, the number of detections provided by ALMACAL is competitive with dedicated ALMA cosmological surveys, offering a complementary way to refine the counts.
We also report the detected numbers of synchrotron sources in Table~\ref{tab:statistics}.
In this paper, we mainly focus on the number counts of DSFGs.

\begin{figure*}
  \centering
  \includegraphics[width=.48\linewidth]{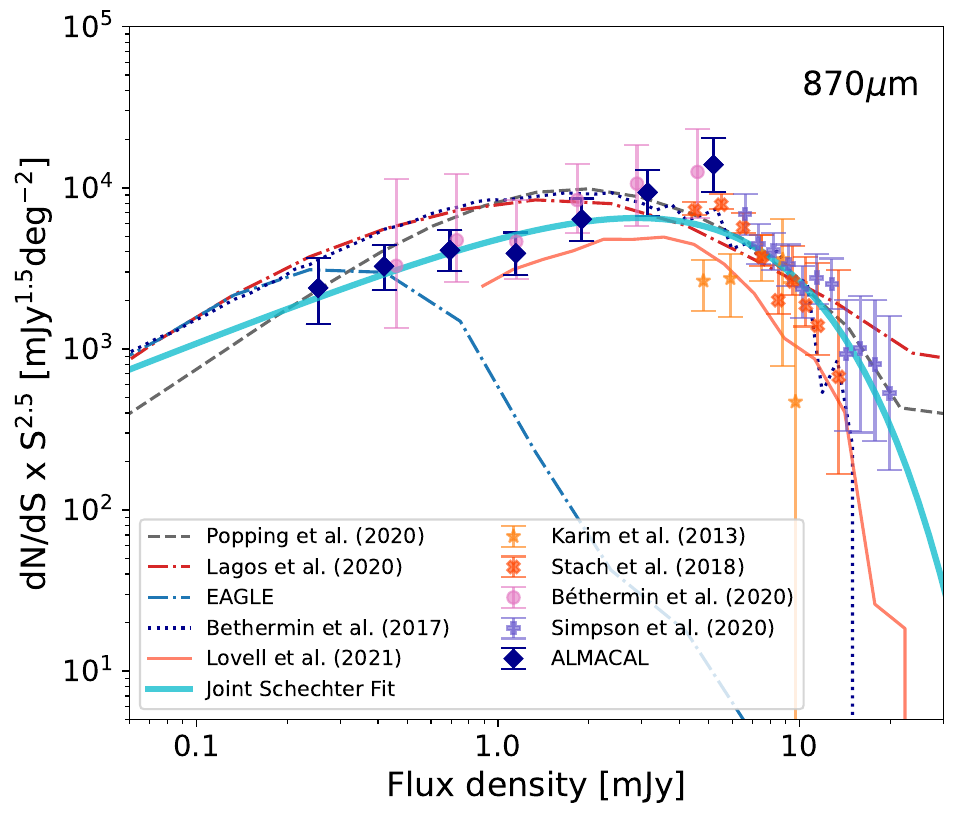}
  \includegraphics[width=.48\linewidth]{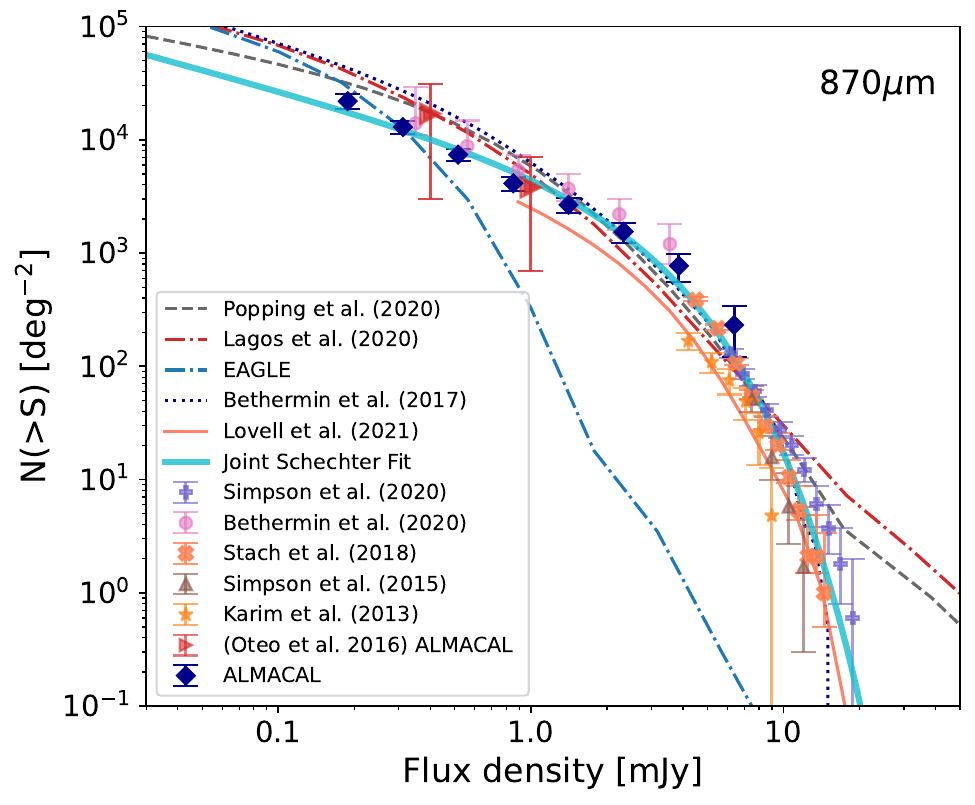}
  \caption{Differential number counts (left) and cumulative number counts (right) at 870\,$\mu$m in band 7. The differential counts have been normalised by $S^{2.5}$ to reduce the dynamic range. We show the new ALMACAL results and the previous ALMACAL band 7 number counts from \citet{Oteo2016}. We also plot the number counts from surveys that used ALMA \citep{Karim2013,Simpson2015a,Stach2018,Bethermin2020,Simpson2020}. Model predictions from Fig.~\ref{fig:number_counts_all} are included here, with the same line styles. The predictions based on EAGLE \citep{Camps2018,McAlpine2019} and SIMBA \citep{Lovell2021} cosmological hydrodynamic simulations are also included as a comparison. The radiative transfer post-processing is only applied to SIMBA galaxies with SFR\,$>20\,\text{M}_\odot\,\text{yr}^{-1}$, which can only give complete number counts above $S_{\rm 870\mu m}>$1mJy. The number counts from ALMACAL are consistent with previous results in a wide flux density range and represent the deepest survey available at the moment. The best joint Schechter fits are shown in each plot.}
  \label{fig:number_counts_B7}
\end{figure*}

\subsection{DSFG number counts}\label{sec:number_counts}

One of the major goals of ALMACAL is to constrain the number counts of DSFGs.
The multi-band coverage of the calibrators makes it possible to constrain the number counts at multiple wavelengthes.

We derive the number counts following the formula used in \citet{Oteo2016}.
For every individual detection, its contribution to the total cumulative number counts is:
\begin{equation}
  N_i(S_i)= \frac{1-f_{\mathrm{sp}(S_i)}}{C(S_i)\cdot A(S_i)},
\end{equation}
where $S_i$ is the flux density of the detection; $f_{\mathrm sp}(S_i)$ is the fraction of spurious sources at $S_i$, which is equal to unity with our 5$\sigma$ detection threshold; $A(S_i)$ and $C(S_i)$ are the effective area and completeness at $S_i$, respectively.

The final number counts are calculated in flux density bins, $S_j$, evenly spaced in log scale. 
For each bin, the cumulative number counts is defined as:
\begin{equation}
  N_j(>S_j) = \sum^{n_i}_{i} N_i(S_i),
  \label{eq:cumulative_nc}
\end{equation}
where $n_i$ is the number of $N_i$ that satisfies $S_i \leq S_j$. 
The differential number counts for each flux density bin is given by:
\begin{equation}
  \frac{dN_j}{dS} = \frac{\sum^{n_{ij}}_{i}N_i}{S_{j+1} - S_{j}},
  \label{eq:differential_nc}
\end{equation}
where $n_{ij}$ is the number of $N_i$ that meets $S_j < S_i \leq S_{j+1}$.

The uncertainties in the number counts are calculated via Monte-Carlo simulations.
Firstly, the flux density of every detection was randomly sampled according to its 1$\sigma$ uncertainty.
Then, the whole DSFG sample is grouped into the different flux density bins, $S_j$.
The flux density bins are constructed to include at least three detections.
Their cumulative and differential number counts are calculated using Equations \ref{eq:cumulative_nc} and \ref{eq:differential_nc}, respectively.
We ran this simulation $1,000$ times to determine the scatter of the number counts.
Finally, Poisson errors were added to the uncertainties of number count according to the number of detections in each bin.
We discuss the number counts that include the unclassified sources in \S\ref{appendixfig:number_counts_perturbation}.

To compare with literature results and theoretical models, we re-scaled the flux densities of all our detections to reference wavelengths in each ALMA band.
The reference wavelengths used in this work are summarised in Table~\ref{tab:summary}.
The re-scaling factor is determined by the flux density ratio at different wavelengths assuming the composite SED from the AS2UDS survey at a reference redshift of $z=2.0$ \citep{Dudzeviciute2020}, which also has a power-law index $\alpha\sim3.2$.
During the scaling, we add 20 per cent of error to the number counts to reflect the variation of the composite SED \citep{Dudzeviciute2020}.
The wavelength re-scaling does not provide a perfect correction for every detection, because each will have a different, unknown redshift, but since the correction is small ($\sim 10$ per cent), it should be reasonable in a statistical sense.
The number counts before and after re-scaling are displayed in Fig.~\ref{appendixfig:number_counts_perturbation}.

Table~2 reports our cumulative and differential number counts and they are plotted in Fig.~\ref{fig:number_counts_all}, which shows the number counts in the different ALMA bands.
This is the first time that we have number counts calculated consistently across several bands.
In the following section, we discuss comparisons with the literature, and with several different model predictions, and the implications of these number counts in the context of the underlying galaxy populations and cosmic infrared background.

\begin{figure*}
  \centering
  \includegraphics[width=.48\linewidth]{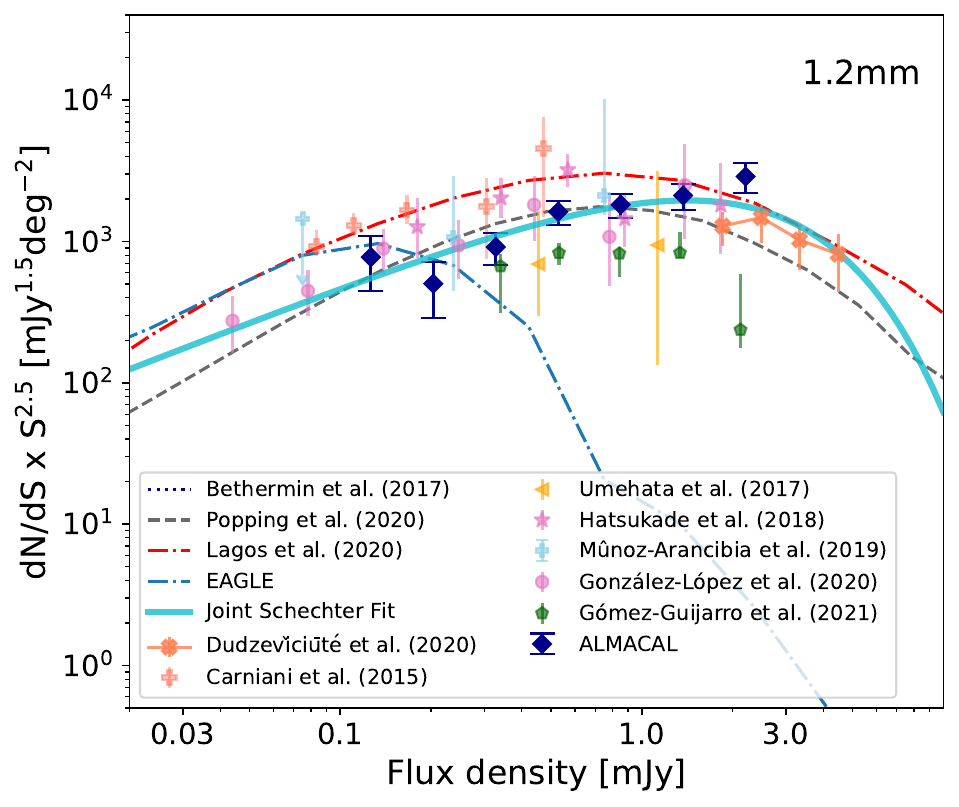}
  \includegraphics[width=.48\linewidth]{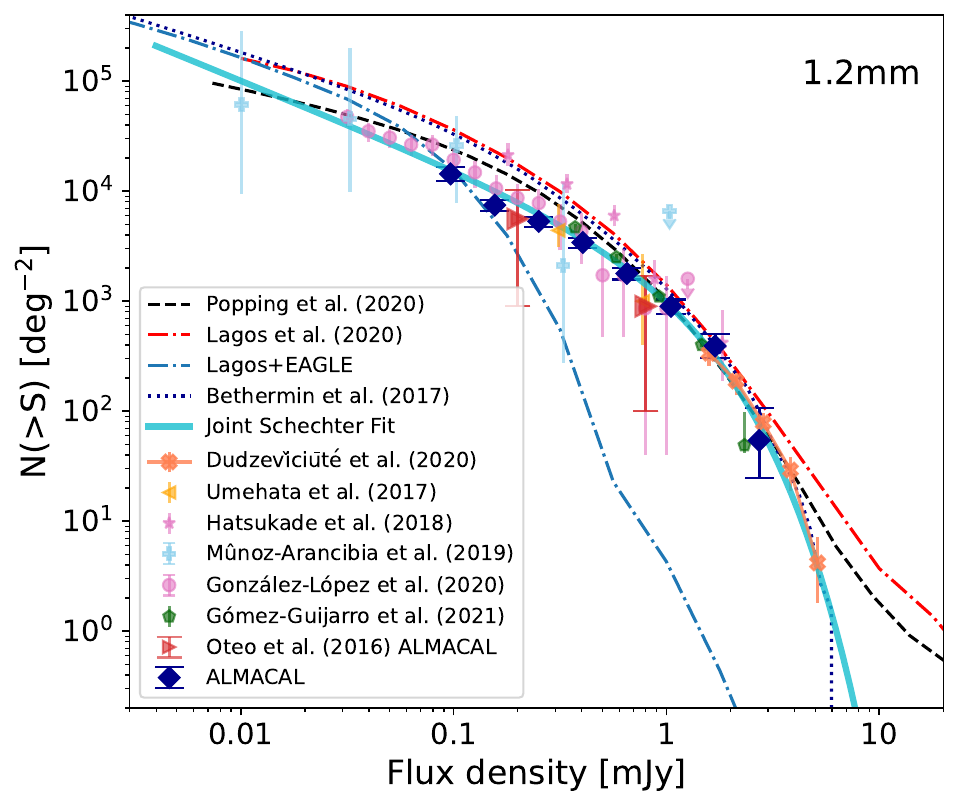}
  \caption{Differential number counts (left) and cumulative number counts (right) at 1.2\,mm. We show the new ALMACAL results and the previous ALMACAL band 6 number counts from \citet{Oteo2016}. Interferometric results reported by \citet{Umehata2018, Hatsukade2018,MunozArancibia2018} are also shown. From \citet{Umehata2018}, only the results from the field are shown. From \citet{MunozArancibia2018}, only the combined results are shown \citep[see the updated results from the corrigendum][]{MunozArancibia2019}. We also show the predicted SED-scaled number counts from AS2UDS \citep{Dudzeviciute2020} to provide constraint in the higher flux density range. Model predictions from Fig.~\ref{fig:number_counts_all} are also included here. In band 6, ALMACAL overlaps with GOODS-ALMA and empirical AS2UDS prediction at the brighter end and is consistent with ASPECS at the fainter end. The best joint Schechter fits are shown in each plot.}
  \label{fig:number_counts_B6}
\end{figure*}
\begin{figure*}
  \centering
  \includegraphics[width=.48\linewidth]{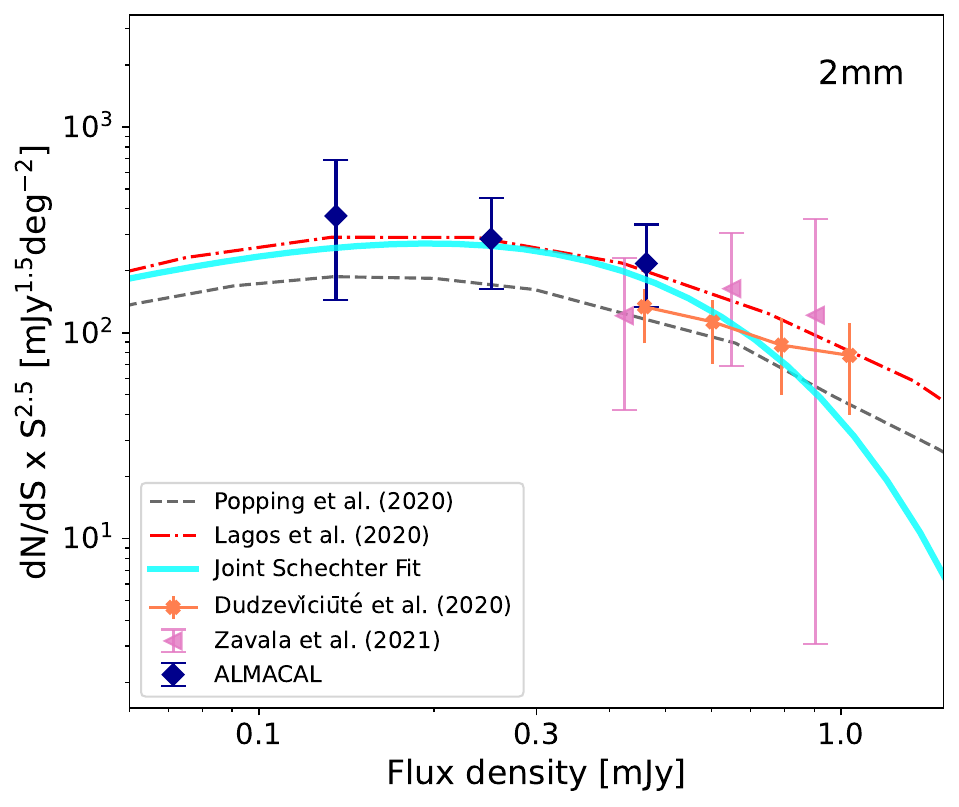}
  \includegraphics[width=.48\linewidth]{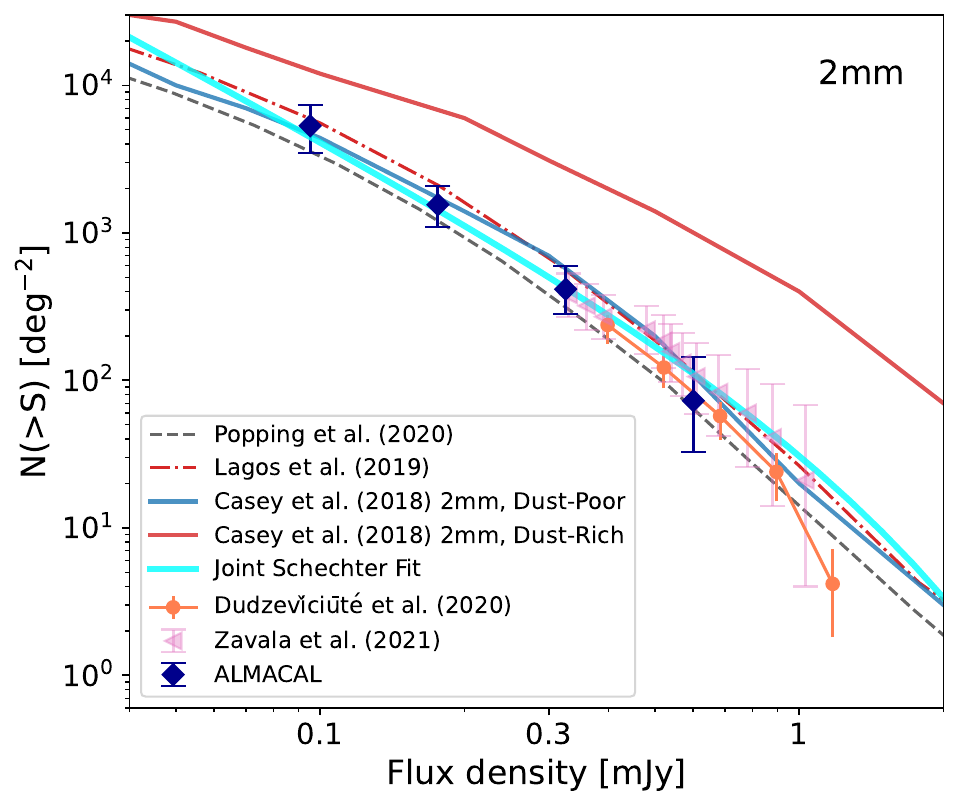}
  \caption{Differential number counts (left) and cumulative number counts (right) at 2\,mm. We show the results from ALMACAL and the Mapping Obscuration to Reionisation with ALMA (MORA) survey \citep{Zavala2021}. We also include the SED predicted 2\,mm number counts from AS2UDS survey \citep{Dudzeviciute2020}. The two different models proposed by \citep{Casey2018a} are both included. Existing data support the dust-poor model and are consistent with the model predictions from \citet{Popping2020} and \citet{Lagos2020}. The best joint Schechter fits are shown in each plot. ALMACAL is consistent with \citet{Zavala2021} and empirical SED-scaled AS2UDS counts at the bright end and present the deepest survey at 2\,mm.}
  \label{fig:number_counts_B4}
\end{figure*}
\begin{figure*}
  \centering
  \includegraphics[width=.48\linewidth]{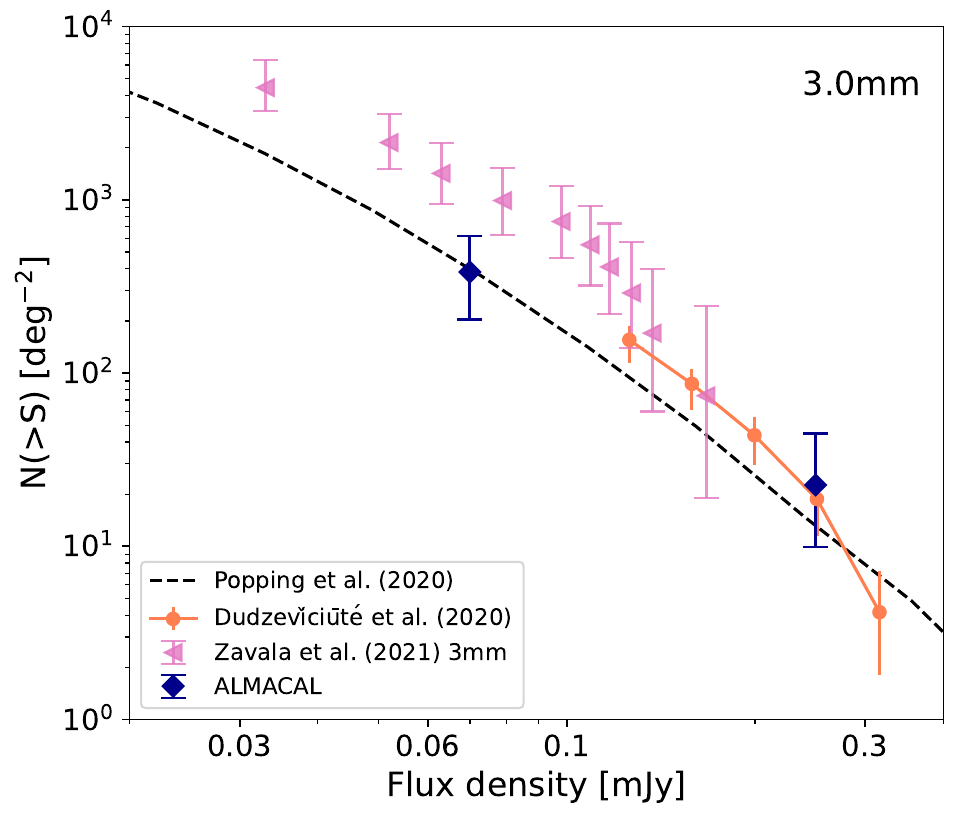}
  \includegraphics[width=.48\linewidth]{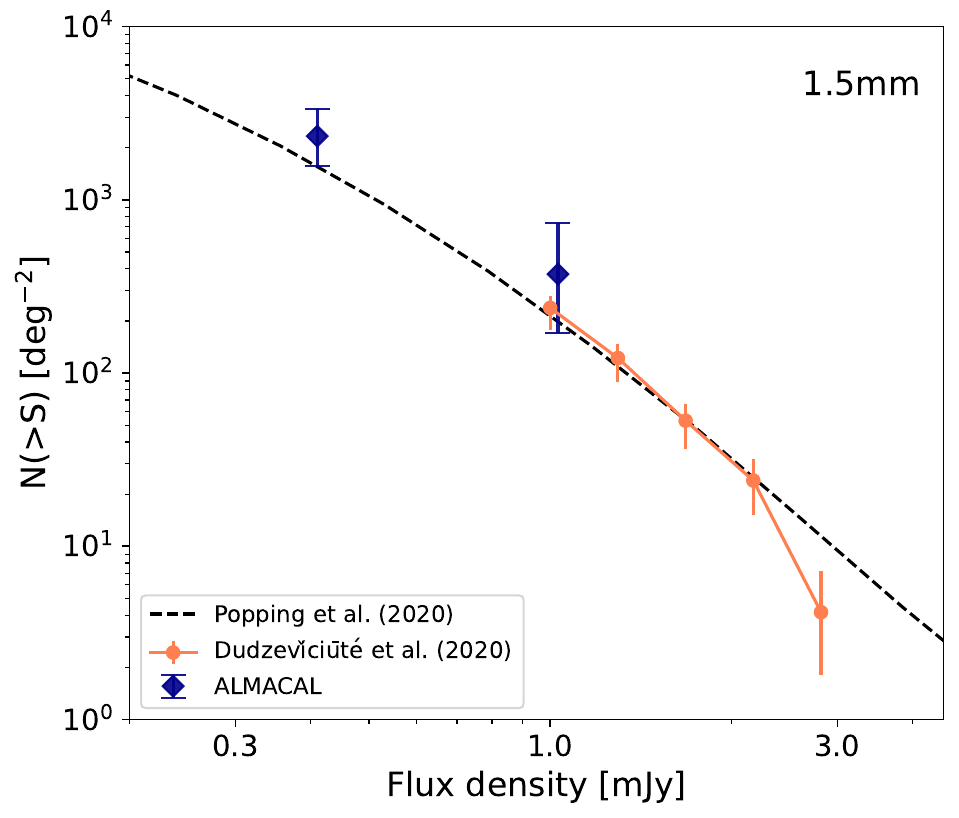}
  \caption{Cumulative number counts at 3\,mm and 1.5\,mm. We also include semi-empirical model predictions from \citet{Popping2020} and the SED-based prediction from AS2UDS survey \citep{Dudzeviciute2020}. At 1.5\,mm, ALMACAL is the first survey to constrain the number counts of DSFGs. At 3.0\,mm, we also include the measurements from \citet{Zavala2021}, which is based on ALMA archival data. Number counts from ALMACAL overlap with \citep{Zavala2021} and empirical SED-scaled AS2UDS predictions at brighter end, but are systematically lower than \citet{Zavala2021} at the faint end.}
  \label{fig:number_counts_B3B5}
\end{figure*}

\begin{table*}
  \caption{Cumulative and differential number counts in various ALMA bands.}
    \makebox[\textwidth][c]{
    \begin{tabular}{ccccccccccccc}
    \toprule
\multicolumn{11}{c}{Band 3 (3\,mm)}\\

\multicolumn{5}{c}{Cumulative Number Counts} & & \multicolumn{5}{c}{Differential Number Counts} \\
\cmidrule(lr){1-5}\cmidrule(lr){7-11}
$S$\,[mJy]  & N & N($>S$)\,[deg$^{-2}$] & $\delta$N($>S$)$_{\rm lower}$ & $\delta$N($>S$)$_{\rm upper}$ & & 
$S$\,[mJy]  & N & dN/d$S$\,[mJy$^{-1}$deg$^{-2}$] & $\delta$(dN/d$S$)$_{\rm lower}$ & $\delta$(dN/d$S$)$_{\rm upper}$ \\ \midrule

0.07 & 7 & 380 & 170& 230 & & 0.17 & 4 & 1990 & 1130 & 1690\\
0.26 & 3 & 20 & 10& 20 & &  $--$ & $--$ & $--$ & $--$ & $--$ \\
\hline
\multicolumn{11}{c}{Band 4 (2\,mm)}\\

\multicolumn{5}{c}{Cumulative Number Counts} & & \multicolumn{5}{c}{Differential Number Counts} \\
\cmidrule(lr){1-5}\cmidrule(lr){7-11}
$S$\,[mJy]  & N & N($>S$)\,[deg$^{-2}$] & $\delta$N($>S$)$_{\rm lower}$ & $\delta$N($>S$)$_{\rm upper}$ & & 
$S$\,[mJy]  & N & dN/d$S$\,[mJy$^{-1}$deg$^{-2}$] & $\delta$(dN/d$S$)$_{\rm lower}$ & $\delta$(dN/d$S$)$_{\rm upper}$ \\ \midrule

0.10 & 21 & 5310 & 1850& 2040 & & 0.14 & 4 & 54440 & 33170 & 47700\\
0.18 & 17 & 1550 & 450& 530 & & 0.25 & 7 & 9050 & 3880 & 5260\\
0.33 & 10 & 410 & 130& 180 & & 0.46 & 7 & 1480 & 560 & 810\\
0.60 & 3 & 70 & 40& 70 & &  $--$ & $--$ & $--$ & $--$ & $--$ \\
\hline
\multicolumn{11}{c}{Band 5 (1.5\,mm)}\\

\multicolumn{5}{c}{Cumulative Number Counts} & & \multicolumn{5}{c}{Differential Number Counts} \\
\cmidrule(lr){1-5}\cmidrule(lr){7-11}
$S$\,[mJy]  & N & N($>S$)\,[deg$^{-2}$] & $\delta$N($>S$)$_{\rm lower}$ & $\delta$N($>S$)$_{\rm upper}$ & & 
$S$\,[mJy]  & N & dN/d$S$\,[mJy$^{-1}$deg$^{-2}$] & $\delta$(dN/d$S$)$_{\rm lower}$ & $\delta$(dN/d$S$)$_{\rm upper}$ \\ \midrule

0.41 & 10 & 2320 & 750& 1020 & & 0.72 & 7 & 4080 & 1570 & 2250\\
1.03 & 3 & 370 & 200& 360 & &  $--$ & $--$ & $--$ & $--$ & $--$ \\
\hline
\multicolumn{11}{c}{Band 6 (1.2\,mm)}\\

\multicolumn{5}{c}{Cumulative Number Counts} & & \multicolumn{5}{c}{Differential Number Counts} \\
\cmidrule(lr){1-5}\cmidrule(lr){7-11}
$S$\,[mJy]  & N & N($>S$)\,[deg$^{-2}$] & $\delta$N($>S$)$_{\rm lower}$ & $\delta$N($>S$)$_{\rm upper}$ & & 
$S$\,[mJy]  & N & dN/d$S$\,[mJy$^{-1}$deg$^{-2}$] & $\delta$(dN/d$S$)$_{\rm lower}$ & $\delta$(dN/d$S$)$_{\rm upper}$ \\ \midrule

0.10 & 132 & 14350 & 2090& 2090 & & 0.13 & 7 & 136600 & 57430 & 78580\\
0.16 & 125 & 7470 & 880& 880 & & 0.20 & 7 & 26900 & 11620 & 15700\\
0.25 & 118 & 5300 & 530& 530 & & 0.33 & 18 & 14820 & 3760 & 4620\\
0.40 & 100 & 3410 & 350& 350 & & 0.53 & 30 & 7990 & 1500 & 1780\\
0.65 & 70 & 1780 & 210& 210 & & 0.85 & 28 & 2720 & 510 & 620\\
1.05 & 42 & 890 & 130& 130 & & 1.37 & 22 & 960 & 200 & 250\\
1.69 & 20 & 390 & 80& 100 & & 2.21 & 17 & 390 & 90 & 120\\
2.73 & 3 & 50 & 20& 50 & &  $--$ & $--$ & $--$ & $--$ & $--$ \\
\hline
\multicolumn{11}{c}{Band 7 (870\,$\mu$m)}\\

\multicolumn{5}{c}{Cumulative Number Counts} & & \multicolumn{5}{c}{Differential Number Counts} \\
\cmidrule(lr){1-5}\cmidrule(lr){7-11}
$S$\,[mJy]  & N & N($>S$)\,[deg$^{-2}$] & $\delta$N($>S$)$_{\rm lower}$ & $\delta$N($>S$)$_{\rm upper}$ & & 
$S$\,[mJy]  & N & dN/d$S$\,[mJy$^{-1}$deg$^{-2}$] & $\delta$(dN/d$S$)$_{\rm lower}$ & $\delta$(dN/d$S$)$_{\rm upper}$ \\ \midrule

0.19 & 92 & 21900 & 3250& 3250 & & 0.25 & 8 & 73860 & 29940 & 39660\\
0.32 & 84 & 12950 & 1640& 1640 & & 0.42 & 15 & 28600 & 8210 & 10180\\
0.52 & 69 & 7390 & 930& 930 & & 0.69 & 16 & 10230 & 2670 & 3350\\
0.86 & 53 & 4120 & 560& 560 & & 1.15 & 14 & 2780 & 740 & 960\\
1.43 & 39 & 2660 & 420& 420 & & 1.90 & 14 & 1280 & 340 & 440\\
2.36 & 25 & 1540 & 300& 370 & & 3.14 & 12 & 530 & 150 & 200\\
3.91 & 13 & 770 & 210& 270 & & 5.19 & 9 & 220 & 70 & 100\\
6.47 & 4 & 230 & 110& 180 & &  $--$ & $--$ & $--$ & $--$ & $--$ \\
\hline

\bottomrule
\end{tabular}%
}
\label{tab:number_counts}%
\end{table*}%

\section{Discussion}\label{sec:discussion}
\subsection{Comparison with the literature}

We compare our number counts reported in Table.~\ref{tab:number_counts} only with those from other interferometric studies
-- all of them from ALMA -- which are relatively free from source blending.
In the following subsections, we will discuss number counts in each band separately, comparing ALMACAL with the literature.

\subsubsection{$870\,\mu$m}
Number counts around $870\,\mu$m -- ALMA band 7 -- were the first to be established in the field of submm cosmology, where useful
data have been collected since the first single-dish survey \citep{Smail1997,Hughes1998,Barger1998,Blain1999}.
In Fig.~\ref{fig:number_counts_B7} we compare the number counts from ALMACAL
with the results from previous ALMA surveys at $870\,\mu$m.
\citet{Karim2013} published the first ALMA interferometric number counts at $870\,\mu$m,
following up the bright single-dish sources which had been detected in the Extended {\it Chandra} Deep Field South using LABOCA on the 12-m APEX telescope \citep[LESS:][]{Weiss2009}.
After that, \citet{Simpson2015a}, \citet{Stach2018} and \citet{Simpson2020} targeted a larger sample
of similarly bright SCUBA-2 sources with ALMA.
More recently, \citet{Bethermin2020} probed fainter flux densities using serendipitous detections from the ALMA Large Programme to INvestigate [C\,{\sc ii}] at Early times (ALPINE) survey. 
Note that this latter survey is expected to be more biased than ALMACAL, since they targeted known high-redshift sources.
With ALMACAL we are going even deeper, detecting sources down to 0.2\,mJy at the 5$\sigma$ level, whilst remaining able to detect relatively bright sources due to the large number of fields covered.
As Fig.~\ref{fig:number_counts_B7} show at the faint end, our number counts follow the general trend seen in earlier work and are consistent with the ALPINE survey \citet{Bethermin2020}.
At the bright end, our results match well with the follow-up ALMA surveys of known, bright submm sources \citep{Karim2013,Stach2018,Simpson2020}.

\subsubsection{1.2\,mm}\label{subsec:number_counts_1.2mm}
Number counts around 1.2\,mm -- ALMA band 6 -- have also been well studied for many years.
It started with single-dish surveys, such as the early single-dish surveys at the IRAM 30-m telescope using the Max-Planck Millimeter Bolometer array (MAMBO) \citep[e.g.][]{Greve2004}, the Bolocam at the 10-m Caltech Submillimeter Observatory \citep[e.g.][]{Laurent2005}, and AzTEC at the 15-m James Clerk Maxwell Telescope and the 10-m Atacama Submillimetre Telescope Experiment (ASTE) \citep[e.g.][]{Scott2012}.
The first deep interferometric observations were made by \citet{Hatsukade2013} 
towards the Subaru/{\it XMM-Newton} Deep Survey field.
Since then, band 6 number counts have been extended 
by mining existing deep fields in the ALMA archive \citep{Ono2014,Carniani2015,Fujimoto2016}.
More recently, dedicated ALMA blind surveys,
ASPECS \citep{Gonzalez-Lopez2019,Gonzalez-Lopez2020} and the ALMA Frontier Fields survey \citep{Gonzalez-Lopez2017,MunozArancibia2018} have greatly improved the depth, while GOODS-ALMA \citep{Franco2018,Gomez-Guijarro2022} has increased the sky coverage substantially, up to 72-arcmin$^2$.
All of these surveys are plotted in Fig.~\ref{fig:number_counts_B6}.
ALMACAL offers a good balance between survey depth and sky coverage.
In the flux density range between 0.1\,mJy and 1.0\,mJy,
most of the number counts are consistent within $2\sigma$. 
At the brighter end, ALMACAL is consistent with GOODS-ALMA and represents the widest survey available.

At 1.2\,mm, discrepancies have been reported in number counts both at the fainter end ($<\!0.1$\,mJy) and at the brighter end ($>1.0$\,mJy).
At the fainter end, \citet{Fujimoto2016} reported number counts that appear systematically high relative to other surveys, possibly due to the use of a relatively low SNR detection threshold (SNR=3$\sigma$) and the bias introduced using the targeted fields.
At these flux limites, we find that ALMACAL is consistent with the results from ASPECS \citep{Gonzalez-Lopez2020} and the ALMA Frontier Fields Survey \citep{Gonzalez-Lopez2017,MunozArancibia2019}.
At the bright end, there are no direct number counts measurements from ALMA for sources brighter than $S_\text{1.2\.mm}\!>$\,2\,mJy.
To better constrain the 1.2\,mm bright-end number counts, we adopted the SED-scaled number counts from AS2UDS \citep{Stach2018,Stach2019}.
\citet{Dudzeviciute2020} derived SED fits for all SMGs within the $\approx$1\,square degree AS2UDS, providing the opportunity to derive the expected counts in other wavelengths.
We thus predict 1.2\,mm flux densities for each SMG based on its best-fitting SED and then derive the number counts after correcting for the original survey completeness down to $S_{\rm 1.2\,mm}\approx 3.6$\,mJy \citep[see also][]{Stach2018}. 
The AS2UDS prediction at 1.2\,mm is over-plotted in Fig.~\ref{fig:number_counts_B6}.
ALMACAL is also consistent with the AS2UDS predictions in the overlapping regime.
In the future, a larger blind ALMA survey, or follow-up of the bright sources from wide field 1.2-mm single-dish surveys, is needed to verify our results and confirm the number counts at the brightest flux densities.

\subsubsection{2\,mm}

There are relatively few published studies of number counts at longer wavelengths $\lambda\gg$\,1\,mm.
The first single-dish 2\,mm survey was conducted by the Goddard IRAM Superconducting Millimeter Observer (GISMO) at the IRAM 30-m telescope \citep{Staguhn2014,Magnelli2019}.
The number counts were later confirmed by the Mapping Obscuration to Reionisation with ALMA (MORA) survey \citep{Zavala2021}.
We revisit the 2\,mm number counts with the detections from ALMACAL.
In Fig.~\ref{fig:number_counts_B4}, we plot the new measurements from ALMACAL, as well as the data from \citet{Magnelli2019} and \citet{Zavala2021}.
Our new measurements are consistent with the two previous surveys at high flux densities and go $4\times$ deeper, down to $S_{\rm 2\,mm}\!\sim$\,0.1\,mJy.
We also predict the 2\,mm number counts based on the AS2UDS sample, following the same method we used for 1.2\,mm in \S\ref{subsec:number_counts_1.2mm}.
The AS2UDS predictions are shown in Fig.~\ref{fig:number_counts_B4}, which matches well with MORA, suggesting that there is no substantial new population of sources appearing at $\lambda\gg$\,1\,mm.

\subsubsection{1.5\,mm and 3\,mm}
ALMACAL is the first survey to constrain the number counts of DSFGs at 1.5\,mm, with detections down to $S_{\rm 1.5\,mm}\!\sim$\,0.4\,mJy.
At 3\,mm, \citet{Zavala2018} provided constraints on the DSFGs number counts using the ALMA Science Archive towards regions around targeted sources in three extragalactic legacy fields: COSMOS, CDF-S and the UDS.
These results were later updated in \citet{Zavala2021} after removed three spurious detections.
We compare our cumulative number counts with those of \citet{Zavala2021} in Fig.~\ref{fig:number_counts_B3B5}, along with the SED-based predictions from AS2UDS and various theoretical model predictions \citep{Lagos2020,Popping2020}.
Our counts agree well with the those predicted from AS2UDS and \citet{Zavala2021} at the bright end. 
However, they are systematically lower than \citeauthor{Zavala2018} at the faint end, which we attribute to their counts being biased high due to associations and potentially source blending in the original target selection.

\begin{table*}
  \caption{Best-fitting models for cumulative and differential number counts.}
    \makebox[\textwidth][c]{
    \begin{tabular}{lcccccccc}
\multicolumn{9}{c}{Cumulative number counts}\\
 \toprule
& \multicolumn{3}{c}{Schechter} & & \multicolumn{4}{c}{Double power-law} \\
    \cmidrule(lr){2-4}\cmidrule(lr){6-9}
       & $\alpha$ & $S_0$ & $N_0$ & & $\alpha_1$ & $\alpha_2$ & $S_0$ & $N_0$ \\ 
    &     &  mJy & deg$^{-2}$  & &     & & mJy & deg$^{-2}$ \\ \midrule 
B4 (2\,mm) & $-1.7_{-0.3}^{+0.3}$& $1.0_{-0.3}^{+0.6}$& $0.09_{-0.01}^{+0.01}$ & & $0.0$ & $2.1_{-0.2}^{+0.2}$ & $1.0$ & $0.04_{-0.01}^{+0.01}$\\
\hline

B6 (1.2\,mm) & $-0.8_{-0.1}^{+0.1}$& $1.0_{-0.1}^{+0.1}$& $2.7_{-0.3}^{+0.3}$ & & $1.0_{-0.1}^{+0.1}$ & $3.8_{-0.3}^{+0.4}$ & $1.2_{-0.2}^{+0.2}$ & $1.6_{-0.2}^{+0.2}$\\
\hline

B7 (870\,$\mu$m) & $-0.6_{-0.1}^{+0.1}$& $2.2_{-0.1}^{+0.1}$& $9.6_{-0.9}^{+0.9}$ & & $1.0_{-0.1}^{+0.1}$ & $4.7_{-0.2}^{+0.2}$ & $4.5_{-0.4}^{+0.4}$ & $3.6_{-0.6}^{+0.7}$\\
\hline

\\

\multicolumn{9}{c}{Differential number counts}\\
\toprule
& \multicolumn{3}{c}{Schechter} & & \multicolumn{4}{c}{Double power-law} \\
    \cmidrule(lr){2-4}\cmidrule(lr){6-9}
    & $\alpha$ & $S_0$ & $N_0$ & & $\alpha_1$  & $\alpha_2$ & $S_0$ & $N_0$  \\ 
    &  &mJy& ${\rm mJy}^{-1}{\rm deg}^{-2}$ & & & & mJy&${\rm mJy}^{-1}{\rm deg}^{-2}$ \\ \midrule
B4 (2\,mm)& $-1.7$ & $0.2_{-0.3}^{+0.4}$& $6.0_{-0.7}^{+0.9}$ & & $0.0$ & $3.1_{-1.1}^{+0.8}$ & $1.0$ & $0.1_{-0.1}^{+0.2}$\\
\hline

B6 (1.2\,mm)& $-1.7_{-0.1}^{+0.1}$& $1.7_{-0.3}^{+0.5}$& $2.4_{-0.9}^{+0.8}$ & & $1.9_{-0.2}^{+0.1}$ & $4.6_{-1.0}^{+1.8}$ & $2.3_{-0.9}^{+0.8}$ & $1.1_{-0.5}^{+1.6}$\\
\hline

B7 (870\,$\mu$m)& $-1.7_{-0.2}^{+0.2}$& $3.9_{-0.7}^{+0.7}$& $2.2_{-0.9}^{+1.4}$ & & $2.0_{-0.3}^{+0.2}$ & $4.7_{-0.5}^{+0.9}$ & $5.6_{-1.5}^{+2.2}$ & $0.8_{-0.8}^{+1.6}$\\
\hline
\\

\multicolumn{9}{c}{Joint fit}\\
 \toprule
  & $\alpha$/$\alpha_1$ & /$\alpha_2$
       & $S_{\rm 2.0\,mm}$ & $N_{\rm 2.0\,mm}$ 
       & $S_{\rm 1.2\,mm}$ & $N_{\rm 1.2\,mm}$
       & $S_{\rm 870\,\mu m}$ & $N_{\rm 870\,\mu m}$\\  
    &  &   
    & mJy & ${\rm mJy}^{-1}{\rm deg}^{-2}$ 
    & mJy & ${\rm mJy}^{-1}{\rm deg}^{-2}$
    & mJy & ${\rm mJy}^{-1}{\rm deg}^{-2}$ \\
    \midrule 
 Schechter  & $-1.7_{-0.1}^{+0.1}$ &
 & $0.3_{-0.1}^{+0.1}$ & $5.0_{-4.0}^{+6.9}$
 & $1.9_{-0.3}^{+0.5}$ & $1.8_{-0.7}^{+0.8}$ 
 & $3.9_{-0.5}^{+0.5}$ & $2.1_{-0.5}^{+0.7}$ \\

\hline

 Double power-law  & $1.9_{-0.1}^{+0.1}$ & $4.2_{-0.3}^{+0.3}$
 & $0.3_{-0.1}^{+0.1}$ & $3.7_{-3.9}^{+6.9}$
 & $2.2_{-0.4}^{+0.5}$ & $1.1_{-0.5}^{+0.7}$ 
 & $4.3_{-0.6}^{+0.7}$ & $1.3_{-0.5}^{+0.6}$ \\
 
\bottomrule
\end{tabular}%
}
\flushleft
  \emph{Note}: Values without errors are fixed values during the fitting.
\label{tab:number_counts_fitting}%
\end{table*}%

\subsubsection{Joint fitting}\label{subsec:joint_fitting}
Combining different surveys offers a way to mitigate some of the biases that are particular to each individual survey.
We have therefore perform a joint fit of the count data in ALMA bands where data is sufficient: 870\,$\mu$m, 1.2\,mm abd 2\,mm.
In band 7 (870\,$\mu$m), we include the data from \citet{Karim2013,Simpson2015a,Stach2018,Bethermin2020,Simpson2020};
in band 6 (1.2\,mm), we include the data from \citet{Umehata2017,MunozArancibia2018,Gonzalez-Lopez2020} and the SED-scaled number counts from AS2UDS \citep{Stach2018};
in band 4, we include the data from \citet{Zavala2021}.
We fit the differential and the cumulative number counts with the Schechter function \citep{Schechter1976} and double power law:
\begin{equation}
  N(>\!S)\,dS = N_0\left(\frac{S}{S_0}\right)^{\alpha}\exp{\left(-\frac{S}{S_0}\right)}\,d\left(\frac{S}{S_0}\right) \;\;,
  \label{eq:Schechter}
\end{equation}
\begin{equation}
  N(>\!S)\,dS = N_0\left[\left(\frac{S}{S_0}\right)^{\alpha_1} + \left(\frac{S}{S_0}\right)^{\alpha_2}\right]^{-1}\,d\left(\frac{S}{S_0}\right) \;\;,
  \label{eq:double_powerlaw}
\end{equation}

\noindent
where $N_0$ is the normalisation factor, in units of deg$^{-2}$ for the cumulative number counts and in units of ${\rm mJy}^{-1}{\rm deg}^{-2}$ for the differential number counts.
$S_0$ is the flux density at the turnover.
The index, $\alpha$, is the power-law index of the Schechter function in Equation~\ref{eq:Schechter};
while $\alpha_1$ and $\alpha_2$ are the slopes of the two independent power laws in Equation~\ref{eq:double_powerlaw}.
During the fitting, $N_0$, $S_0$ and $\alpha$, or $\alpha_1$ and $\alpha_2$ are free parameters.
The best-fit models for the different bands are summarised in Table~\ref{tab:number_counts_fitting} and the best-fit Schechter functions are plotted in Figs~\ref{fig:number_counts_B7}, \ref{fig:number_counts_B6} and \ref{fig:number_counts_B4}.
We adopt the best-fit Schechter function as our fiducial model, because it provides the best fit to the flattening trend in the very deep 1.2-mm number counts (see discussion in \citealt{Gonzalez-Lopez2020}).

We constrain the free parameters using the maximum-likelihood minimisation algorithm, {\sc minimize}, from {\sc scipy.optimize}, and adopt the Markov chain Monte Carlo (MCMC) sampler, {\sc emcee}, to derive the confidence levels.
We treat different surveys equally, using their own reported errors, but increase the uncertrainties if re-scaling is applied (see more in \S\ref{sec:number_counts}).
In band 4, the available data are not sufficient to constrain the models. 
We thus fix $\alpha=1.7$ for the Schechter function and $\alpha_1=0$ and $S_0=1$\,mJy for the double power law (see more discussion in \S\ref{subsec:galaxy_populations}).

From our best fits, we confirm the shallower trend seen in the faint flux density range at 870\,$\mu$m \citep{Bethermin2020} and 1.2\,mm \citep{Gonzalez-Lopez2020}.
Based on the joint Schechter function fitting for the differential number counts, the turnover flux density at 870\,$\mu$m is $S_{870\mu{\rm m}}=3.9_{-0.7}^{+0.7}$\,mJy.
At 1.2\,mm, \citet{Gonzalez-Lopez2020} found that the number counts flattened below 0.1\,mJy, and argued for a triple power law to fit the differential number counts, based on their $P(D)$ analysis.
In our fitting, a Schechter function with a turnover at $S_{1.2{\rm mm}}=1.7_{-0.3}^{+0.5}$\,mJy with a power-law index of $\alpha=1.7\pm0.1$ gives a reasonable fit to the existing data. 
However, as we mentioned earlier, considerable uncertainties exist both at the fainter and brighter end of the 1.2\,mm number counts.
At 2\,mm, the constrained fitting gives a turnover flux $S_{2{\rm mm}}=0.2_{-0.3}^{+0.4}$\,mJy, which still suffers from large uncertainties.

\begin{figure*}
  \centering
  \includegraphics[width=.48\linewidth]{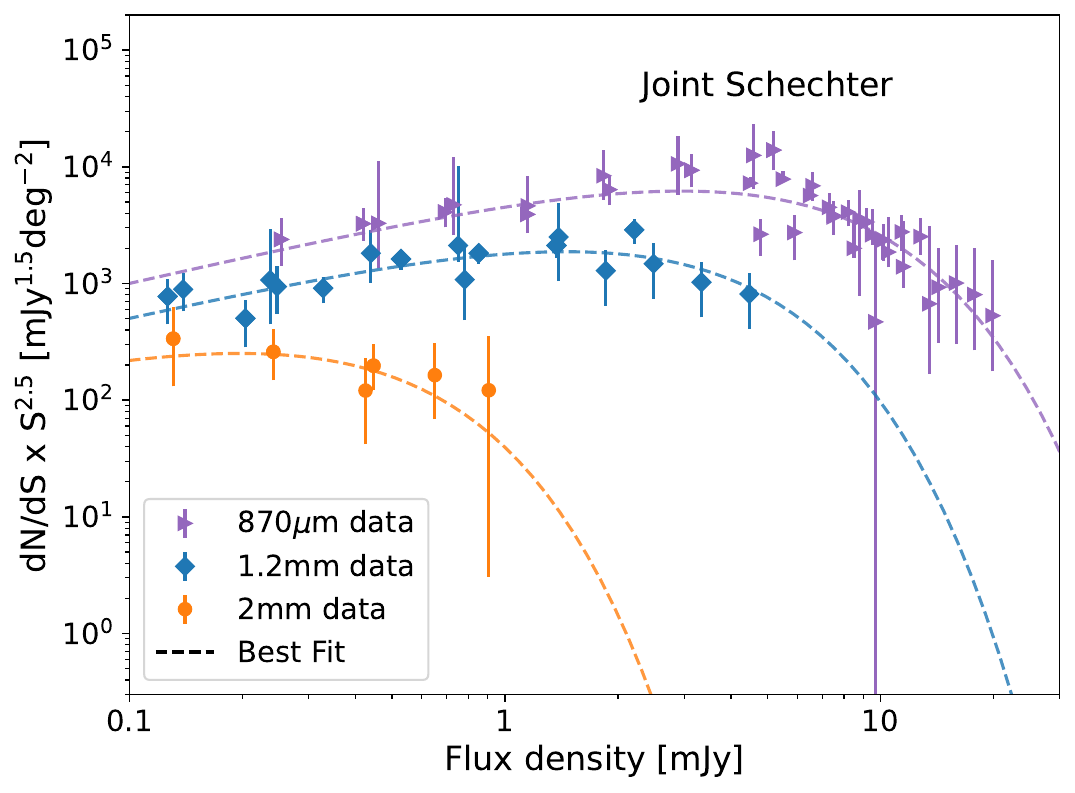}
  \includegraphics[width=.48\linewidth]{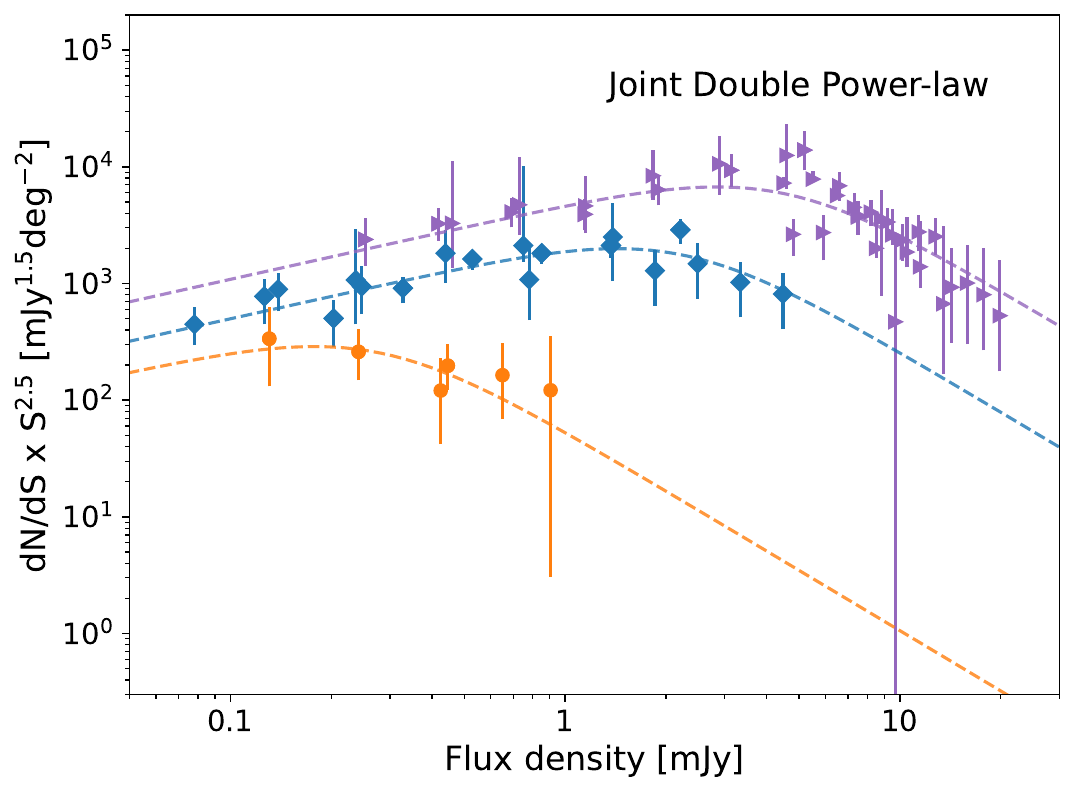}
  \caption{Joint fits combining differential number counts at 2\,mm, 1.2\,mm and 870\,$\mu$m. 
  {\it Left:} Joint Schechter fits for all the data mentioned in \S\ref{subsec:joint_fitting}. During the fitting, the $\alpha$ index of the Schechter function is bound for all the three wavelengths. The best final fits give $\alpha=-1.7\pm0.1$.
  {\it Right:} Joint double power-law fits. In the flux density regime of
ALMACAL, all the measurements can be fitted with the same power-law index $\alpha=-1.9\pm0.1$, indicating the same or similar galaxy population behind the number counts.}
  \label{fig:joint_single_fitting}
\end{figure*}

\subsection{Comparing with model predictions}
\label{subsec:data_model_comparison}

Simulations and analytical models of galaxy formation and evolution must also explain the DSFG population.
Depending on basic assumptions and underlining techniques, these modelling methods can be roughly divided into semi-analytic models (SAMs), semi-empirical models and hydrodynamic (N-body) simulations.

SAMs are the most widely used tools \citep{Cole2000}.
They start with the merging trees of dark matter halo from either cosmological $N$-body simulations or self-consistent Monte-Carlo simulations and use analytical equations for baryonic processes.
Their flexibility and relatively cheap computational cost make them powerful to explore large regions of parameter space.
However, it was difficult to reproduce submm/mm number counts by incorporating standard feedback processes along with radiative transfer to balance the FUV-to-NIR and the FIR-to-submm/mm emission of galaxies simultaneously across cosmic time \citep{Baugh2005,Somerville2012,Lacey2016,Cowley2019}.
\citet{Lagos2019,Lagos2020} adopted attenuation curves obtained from the radiative transfer analysis of hydrodynamical simulations \citep{Trayford2020} to reproduce the panchromatic emission of galaxies.
Their model is better able to match the multi-wavelength DSFG number counts. 
We show predictions from \citet{Lagos2020} in Fig.~\ref{fig:number_counts_all}. 
They are broadly consistent with ALMACAL in the various ALMA bands, but exhibit over-prediction at the fainter end of bands 6 and 7.

Due to the increasing complexity of SAMs and the computational cost of fully three-dimensional dust radiative transfer, observational empirical scaling relations have been coupled into SAMs -- creating semi-empirical models -- to simplify the calculations \citep{Bethermin2017,Popping2020}.
\citet{Popping2020} provide a new framework to explain the number counts at 1.2\,mm \citep{Gonzalez-Lopez2020}, modelling the submm/mm emission of galaxies with simple functions from full radiative transfer simulations, where the submm/mm flux density of a galaxy is a function of SFR and total dust mass \citep{Hayward2011}.
The \citeauthor{Popping2020} model is extended here to provide the predictions on the Rayleigh-Jeans tail of the submm SED by fitting the original 850-$\mu$m and 1.1-mm predictions with a grey body.
Their predictions for different ALMA bands are shown in Fig.~\ref{fig:number_counts_all}.
They are slightly lower than the predictions from \citet{Lagos2020} at fainter flux densities and are more consistent with the observations but at the cost that there are no true ab initio predictions. 
\citet{Casey2018a} also explored the number counts with different fractions of DSFGs at $z>4$. 
They proposed two simple models: their dust-poor model has a steep slope in the luminosity function in the early Universe, while their dust-rich model has a much shallower slope and predicts many more DSFGs.
The two models thus predict different number counts in the mm bands,
with the dust-rich model having a higher number density of DSFGs.
Our 2\,mm number counts favour the dust-poor model, which is also closer to the prediction from \citet{Popping2020} and \citet{Lagos2020}.

Hydrodynamic models, on the other hand, still struggle to reproduce the number counts of DSFGs and usually underpredict their numbers \citep{Shimizu2012}.
With recent advances in computational resources, fully 3-D radiative transfer (RT) has been integrated with modern cosmological simulations with different recipes for stellar and AGN feedback \citep{Camps2018,McAlpine2019,Lovell2021}.
Here, we compare our results with EAGLE (Evolution and Assembly of GaLaxies and their Environment, \citealt{Schaye2015}) and SIMBA \citep{Dave2019} cosmological simulations that have 3-D radiative transfer implementations \citep{Camps2018,Lovell2021}.
To compute the number counts of EAGLE, we first retrieve the publicly available observer-frame fluxes in ALMA bands 6 and 7 of all galaxies from $z=10$ to $z=0$ \citep[see details about the flux calculation in][]{Camps2018}.
Then, we use the fluxes and all the snapshots of the simulation to construct a lightcone. 
For each redshift, we compute the projected sky area of 100 Mpc$^2$ and the redshift bin implied by a 100\,Mpc comoving distance.
We then calculate the number of galaxies per unit area, per unit redshift, at each flux bin contributed by each snapshot $N(z,S)$ .
Finally, We integrate $N(z,S)$ along redshift to obtain the cumulative number counts $N(S)$.
For SIMBA, we adopt the results from \citet{Lovell2021}, which reproduced the population of bright DSFGs but -- due to the computational cost -- their predicted number counts are only complete down to $S_{\rm 850\mu m}\!\sim1$\,mJy.
We show all the predictions in Fig.~\ref{fig:number_counts_B7} and Fig.~\ref{fig:number_counts_B6}.

Modelling the basic observations can help us to understand the relevant physical processes behind galaxy formation and evolution.
Number counts are one of the most basic measurements from observations and play an essential role in validating models.
By construction, semi-empirical models closely match the observed scaling relations, producing reasonable number counts at most of the observable flux densities, which can help design future surveys.
SAMs start with basic physical assumptions, representing our understanding of the physics behind the observables. 
They can already give reasonable predictions for the flux densities covered by existing surveys.
Future surveys with the flexibility offered by SAMs will continue to be a powerful tool to understand the fundamental physics behind the number counts.
Hydrodynamic cosmological simulations remain the most computational expensive method. However, they also offer the most detailed description of the internal/external galactic structures and their environments. 
The discrepancy between the model predictions and the observations will continue to motivate more detailed sub-grid physics in cosmological simulations.
Multi-wavelength number counts -- especially in the submm/mm range -- offer additional constraints on the redshift distribution of DSFGs, thanks to the negative-K correction \citep{Baugh2005,Casey2014}.
Future, deep multi-band DSFG survey will therefore continue to be a critical benchmark for modern galaxy evolution models.

\begin{figure*}
  \centering
  \includegraphics[width=.6\linewidth]{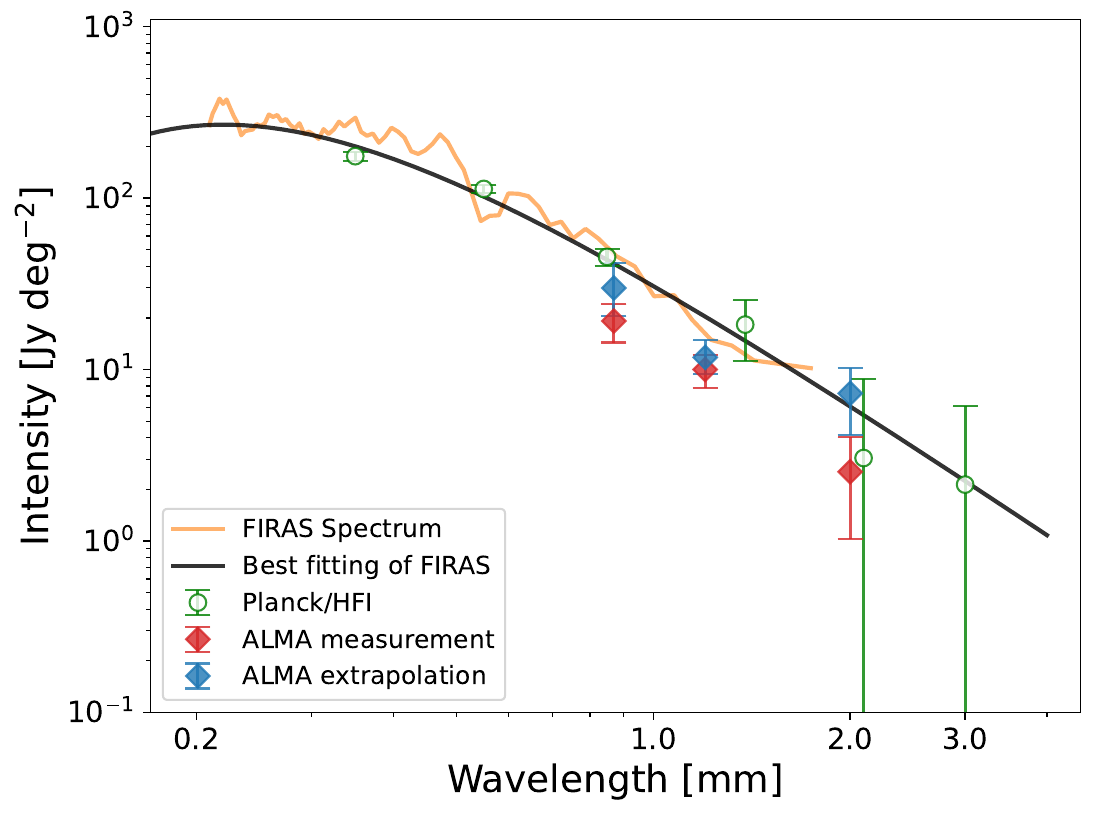}
  \caption{ALMA resolved cosmic infrared background. The resolved CIB at different ALMA bands is the integration of joint Schechter fitted differential number counts within the observed flux density ranges. The predicted CIB is the whole integration of the best-fitting differential number counts. The direct measurement from FIRAS on {\it COBE} is shown in orange. The black curve is the best-fitting FIRAS spectrum made by \citet{Fixsen1998}.
  We also show the recalibrated {\it Planck}/HFI data \citep{Odegard2019}. Comparing with the direct measurements from FIRAS, ALMA currently resolved nearly one half of the CIB from 870$\mu$m to 2\,mm, but the fraction is highly uncertain due to the large uncertainties of the FIRAS spectrum at the submm/mm wavelengths.}
  \label{fig:resolved_CIB}
\end{figure*}
\begin{table*}
  \centering
  \caption{ALMA resolved cosmic infrared background.}
  \label{tab:ALMA_CIB}
  \begin{tabular}{ccccc}
    \hline\hline
    Wavelength  & Resolved Intensity & Extrapolated Intensity & Resolved fraction $^a$ & Resolved fraction $^b$ \\ 
    mm & Jy\,${\rm deg}^{-2}$  &  Jy\,${\rm d
eg}^{-2}$ & Observed & Extrapolation\\ 
    \hline
2000 & $2.4\pm1.7$ & $7.0\pm3.6$ & $0.40_{-0.11}^{+0.23}$ & $0.36_{-0.25}^{+0.27}$ \\ \\[-1em]
1200 & $11.2\pm2.2$ & $14.1\pm2.9$ & $0.54_{-0.15}^{+0.32}$ & $0.79_{-0.15}^{+0.18}$ \\ \\[-1em]
870 & $18.3\pm4.7$ & $27.0\pm7.8$ & $0.44_{-0.12}^{+0.26}$ & $0.67_{-0.17}^{+0.17}$ \\ \\[-1em]
    \hline\hline
  \end{tabular}
  \flushleft
  \emph{Note}: The resolved intensity is the integration within the flux density range that is covered by ALMA; a): the resolved fraction compared with the FIRAS measurement; b): the resolved fraction compared with the prediction of this work.
\end{table*}

\subsection{The galaxy populations behind the number counts}\label{subsec:galaxy_populations}
It has long been argued that submm/mm surveys at different depths and different wavelengths probe different galaxy populations and redshifts \citep{Chapman2005,Casey2014}.
This claim has been explored by semi-empirical models and semi-analytical models \citep{Bethermin2015, Popping2020, Lagos2020}.
\citet{Bethermin2015} argued that shallower surveys at longer wavelengths yield DSFGs with a higher median redshift. 
\citet{Popping2020} emphasised that the contribution from galaxies at $z>4$ to the number counts is small and the flattening of the number counts at $S_{\rm 1.2mm}<0.1$ is likely caused by the shallow faint-end luminosity functions of $z=1$--2 DSFGs.
\citet{Lagos2020} also suggested that the redshift increases with flux density for bright SMGs, but this trend flattens out for fainter DSFGs.
In general, simulations reproduced the observational trend that reshift increases with flux density for bright SMGs \citep[][see also \citealt{Ivison2007,Simpson2017,Brisbin2017,Stach2019}]{Simpson2020}, but we still have limited observational constraints about the redshift distribution of fainter DSFGs.

Thanks to the unique wavelength coverage of ALMACAL, we have been able to constrain the multi-band number counts for faint DSFGs simultaneously.
Even though all the parameters are fitted independently, band 6 and 7 share remarkable similar power-law indices (see Table~\ref{tab:number_counts_fitting}).
If we were probing a similar galaxy population at different wavelengths and if their redshift distributions do not vary significantly, the conversion of number counts between different wavelengths could then be approximated by a simple colour difference.
Inspired by this idea, we jointly fit the differential number counts in band 6 and band 7 by linking their power-law indices. 
In Fig.~\ref{fig:joint_single_fitting}, we show the joint Schechter and double power-law fitting of ALMACAL measurements.
In the flux density regime of ALMACAL, the differential number counts in band 6 and band 7 can be well fit with the same power-law index with $\alpha_1=-1.9\pm0.1$, implying the number counts in band 6 and band 7 are dominated by the same galaxy population.
A tentative fit with the same power-law index can also be applied to the fainter end number counts at 2\,mm, but the statistical significance is limited.
The joint Schechter fitting gives $\alpha=1.7\pm0.1$, which is the same for independent fitting for band 6 and 7.
To give more robust total CIB prediction in band 4, we fixed $\alpha=1.7$ for the Schechter fitting of differential number counts.

\subsection{Resolved cosmic infrared background}
Two methods have been used to constrain the CIB: the direct measurement and the integrated intensity based on galaxy number counts.
For direct measurements, the pioneering experiment were achieved by the Far Infrared Absolute Spectrophotometer (FIRAS) aboard {\it Cosmic Background Explorer} ({\it COBE}) \citep{Puget1996, Fixsen1998} and were later continued by instruments including {\it ISOPHOT} \citep{Juvela2009}, {\it AKARI} \citep{Matsuura2011} and {\it Planck}/HFI \citep{PlanckCollaboration2014} at different wavebands.
A direct measurement requires accurate foreground modeling for the scattered light from solar interplanetary dust and the emission from the Milky Way, which usually needs additional assumptions or ancillary data for calibration \citep{Fixsen1998,Odegard2019}.
On the other hand, if we have good knowledge of the galaxy number counts, we can also reconstruct the CIB by integrating the intensities from all the galaxies at different wavelengths.
Due to the limiting sensitivity of the instruments used to derive number counts, extrapolations are normally needed to recover the undetected faint sources.
Comparing the integrated energy density of galaxy number counts with direct measurements, we can quantify the energy budgets of different galaxy populations.
The accurate comparison between the two can also provide us clues about the presence of any unidentified populations.

After nearly two decades of efforts, a major part of the CIB has been resolved into single sources \citep{Casey2014}.
However, large uncertainties still exist in the submm/mm bands.
Firstly, the accuracy of direct CIB measurement in the original FIRAS spectrum is decreasing toward longer wavelengths, as one can see in Fig.~\ref{fig:resolved_CIB}.
Secondly, the integration based on galaxy number counts also suffers from large uncertainties due to the poorly constrained slope at faint flux densities andalso cosmic variance uncertainties. 
Because of this, different works report quite different fractions of resolved CIB, both at 870\,$\mu$m \citep[e.g.][]{Smail2002,Chen2013,Oteo2016,Bethermin2020} and 1.2\,mm \citep[e.g.][]{Fujimoto2016,Gonzalez-Lopez2020}.

To make optimal use of the existing data and to minimize their biases, we choose to derive the resolved CIB using our joint best-fitting model.
Two values are calculated: the resolved intensity and the extrapolated total intensity.
The resolved intensity is the integration of the best-fitting differential number counts within the flux density range that has been covered by existing surveys.
The extrapolated total intensity is based on the extrapolation of the best-fitting differential number counts to include both faint and bright sources that are beyond the detection limits of current surveys.
It should be noted that the extrapolation can introduce large uncertainties if we adopt incorrect models.
The deep ALMA blind surveys ASPECS and the ALMA Frontier Fields Survey at 1.2\,mm \citep{Gonzalez-Lopez2020,MunozArancibia2018} suggested that the number counts flatten out at the very faint flux density end ($S_\text{1.2\,mm}\!<$\,0.1mJy), implying that the counts can be best described by a Schechter function. 
We thus made the predictions based on our best-fitting Schechter functions, using the theoretical integration formula:

\begin{equation}
  \int_0^{\infty} N(>\!S)\,S\,dS = N_0\times S_0 \times \Gamma(\alpha+2) \;\;\;.
  \label{eq:Integrated_Schechter}
\end{equation}

\noindent
The resolved CIB and the predictions at 2\,mm, 1.2\,mm, and 870\,$\mu$m are summarized in Table~\ref{tab:ALMA_CIB}.

In Fig.~\ref{fig:resolved_CIB}, we compare our results with the direct measurements from {\it COBE}/FIRAS and the most recent re-calibrated {\it Planck}/HFI data \citep{Odegard2019}.
Both the resolved intensities and the predictions from ALMA are shown.

Our results are consistent with the literature and our predictions are close to the direct measurements from FIRAS.
At 870\,$\mu$m, the resolved CIB for $S_{870\mu\text{m}} > 0.2$\,mJy is $18.3\pm4.7$\cib{}, 
which matches the resolved intensity of $16.4\pm2.7$\cib{} from the secure sample  of \citet{Bethermin2020}. 
Compared with direct measurements from FIRAS, the resolved intensity at 870\,$\mu$m accounts for $\sim$44 per cent of the total CIB.
At 1.2\,mm, our resolved intensity for $S_{\rm 1.2\,mm}>45\,\mu$Jy is $11.2\pm2.2$\cib{} and the extrapolated DSFGs' contribution to the CIB is $14.1\pm2.9$\cib{}.
Both of these values are slightly larger than the measurement of ASPECS \citep{Gonzalez-Lopez2020}, but closer to the direct measurements.
At 2\,mm, the resolved CIB for $S_{\rm 2mm}>0.1$mJy is $2.4\pm1.7$\cib{}, which is close to half of the direct measurement.
However, since large uncertainties also exist in the FIRAS measurement at this wavelength, the comparison the calculation of the resolved fraction is not very accurate.
More recently, \citep{Odegard2019} have improved the accuracy of the CIB direct measurement with Planck/HFI data, but the uncertainties at the longer wavelengths have not significantly improved.
Therefore, to have a more precise comparison, we still need more accurate direct measurements of the total CIB at submm/mm wavelengths.
Comparing our predictions based on extrapolations, we have resolved $36^{+27}_{-25}$, $79^{+18}_{-15}$, and $67^{+17}_{-17}$ per cent of the CIB at 2\,mm, 1.2\,mm, and 870\,$\mu$m, respectively.

\section{Conclusions}\label{sec:conclusions}
ALMACAL is a survey that is exploiting calibration data that are accumulating “for free” with every scheduled ALMA observing project.
Before the observatory shutdown due to Covid-19 in March 2020, ALMACAL had accumulated more than 1,000\,h of observing time and covered 1001 calibrator fields.
The sensitivity, total sky coverage, and wide frequency sampling make ALMACAL a promising dataset to undertake blind surveys for DSFGs.
Within the ALMACAL footprints, we detect 371 sources, including 186 DSFGs confirmed by their spectral indices.
We report the number counts based on these DSFGs for ALMA band 3 to band 7 (wavelengths spanning from 3\,mm to 870\,$\mu$m), which are mostly in agreement with existing surveys at overlapping flux densities.
In band 4 (2\,mm) and 7 (870\,$\mu$m), ALMACAL represents the deepest survey available to this date.
In band 5 (1.5\,mm), ALMACAL is the first survey to be able to constrain the number counts of DSFGs.
Together with the previously reported band 8 results from \citet{Klitsch2020}, we are now able to present  number counts covering almost the entire wavelength range of covered by ALMA, from 0.65 to 3\,mm.

We compare our number counts with various model predictions.
The semi-analytic models from \citet{Lagos2020} and semi-empirical models from \citet{Bethermin2017} and \citet{Popping2020} match the number counts in ALMA bands reasonably well.
Hydrodynamic simulations, although finding it generally harder to reproduce the number counts of DSFGs, have shown promising results from the recent efforts \citep[e.g.][]{McAlpine2019,Lovell2021}.
As demonstrated in this work, multi-wavelength number counts can be a powerful benchmark to validate galaxy formation and evolution models.
Future surveys, including the ongoing ALMACAL, will continue to improve the DSFG number counts by covering a larger sample and a wider flux density range.

We also turn ALMACAL into a cosmological survey to constrain the energy budget from DSFGs to the cosmic infrared background.
We provide joint fits for number counts of DSFGs by combining ALMACAL with literature ALMA surveys.
Compared with the direct measurements from FIRAS/{\it COBE} \citep{Fixsen1998} and HFI/{\it Planck}, we report ALMA has directly resolved $40^{+23}_{-11}$, $54^{+32}_{-15}$, and  $44^{+26}_{-12}$ per cent of CIB at 2\,mm, 1.2\,mm, and 870\,$\mu$m, respectively.
Due to the large uncertainties of direct measurements at submm/mm wavebands, we still suffer from large errors, thus demanding more accurate direct CIB measurements.

The large number of detections in ALMACAL suggests the promising opportunity to conduct a large submm/mm survey with ALMA calibration data.
We only explore the continuum image of these calibrator fields, more treasures are still buried in this database.
In addition, ALMACAL is just a small part of the ALMA archive database.
How to make use of the enormous ALMA archive effectively is challenging, but also a promising way to find more hidden gold.

\section*{Data Availability}
All the data used in the work is publish available from ALMA science archive \hyperlink{ALMA Science Archive}{https://almascience.eso.org/alma-data}. The catalogue is available on \hyperlink{ALMACAL}{https://almacal.wordpress.com} and \hyperlink{Git repository}{https://github.com/cjhang/almacal}.

The code used for this paper is also available online: \url{https://github.com/cjhang/almacal}.

\section*{Acknowledgements}
We greatly thank U. Dudzevi\v ci\=ut\'e for providing the SEDs of the AS2UDS sample and Christopher C. Lovell for providing the simulations results from SIMBA.
This work is funded by the Deutsche Forschungsgemeinschaft (DFG, German Research
Foundation) under Germany's Excellence Strategy -- EXC-2094 --
390783311. 
IRS acknowlege the support from STFC(ST/T000244/1).
AK gratefully acknowledges support from the Independent Research Fund Denmark via grant number DFF 8021-00130.
CL has received funding from the ARC Centre of Excellence for All Sky Astrophysics in 3 Dimensions (ASTRO 3D), through project number CE170100013.

This research made use of:
\href{http://www.numpy.org/}{\texttt{NumPy}}, 
    a fundamental package for scientific computing with Python;\ 
\href{http://matplotlib.org/}{\texttt{Matplotlib}}, 
    a plotting library for Python;\ 
\href{http://www.astropy.org/}{\texttt{Astropy}}, a community-developed 
    core Python package for astronomy;\ 
\href{https://ipython.org}{\texttt{IPython}}, 
    an interactive computing system for Python;\ 
\href{https://photutils.readthedocs.io/en/v0.6}{\texttt{photutils}},
    an affiliated package of \texttt{AstroPy} to provide tools for detecting and performing photometry of astronomical sources (\citealt{Bradley2020a});

\bibliographystyle{mnras} 
\bibliography{number_counts} 

\vspace{2em}
This paper has been typeset from a \TeX/\LaTeX{} file prepared by the author.

All the appendices will be on-line only.

\newpage

\appendix

\section{Flux boosting for sources with different sizes}\label{appdixsec:simulation_fluxboosting}
\begin{figure}
   \centering
   \resizebox{\hsize}{!}{\includegraphics{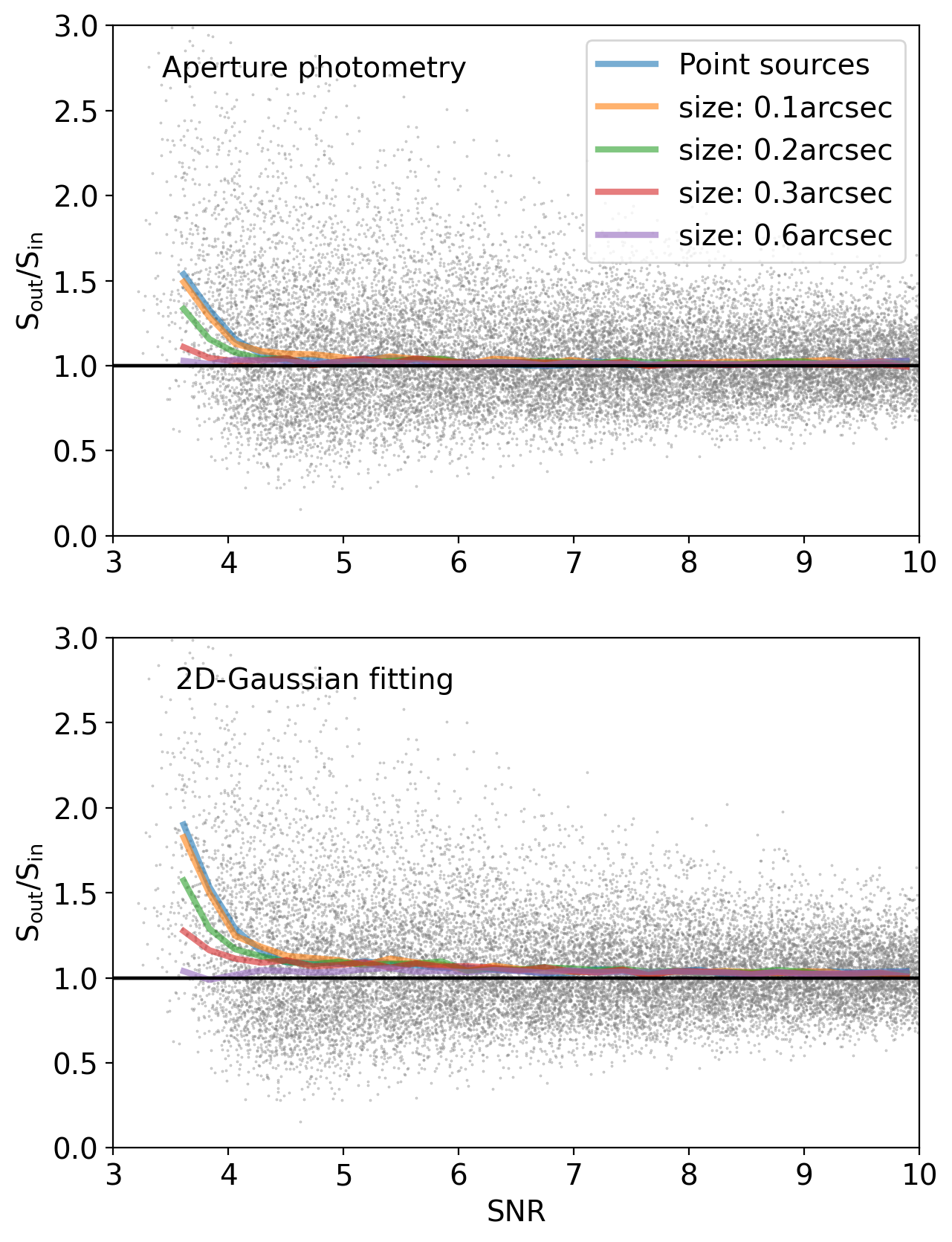}}
   \caption{The comparison of flux boosting for different extended sources. The labels are same as Fig.~\ref{fig:completeness}. Sources with different sizes have different boosting factors but converge at SNR\,$>5$.}%
   \label{appendixfig:flux_boosting_comparison2}
\end{figure}

We use the same simulation as described in \S\ref{sec:flux_boosting} to quantify the flux boosting of different source sizes.
Each time we inject single-size sources and measure their boosting effects.
In Fig.~\ref{appendixfig:flux_boosting_comparison2}, it shows the boosting effects from different methods.
Firstly, sources with different sizes have slightly different boosting factors, especially at low SNRs.
Secondly, despite the very extended sources (FWHM$\sim$0.6\,arcsec), all the boosting factors converged at SNR~$>5$.
Since we only account for sources with SNR~$>5$, it is reasonable to correct the boosting effects based on the simulation of our fiducial source size (FWMH=0.2\,arcsec).

\section{Radial distance and flux densities of all the detections}
\begin{figure}
   \centering
   \includegraphics[width=\linewidth]{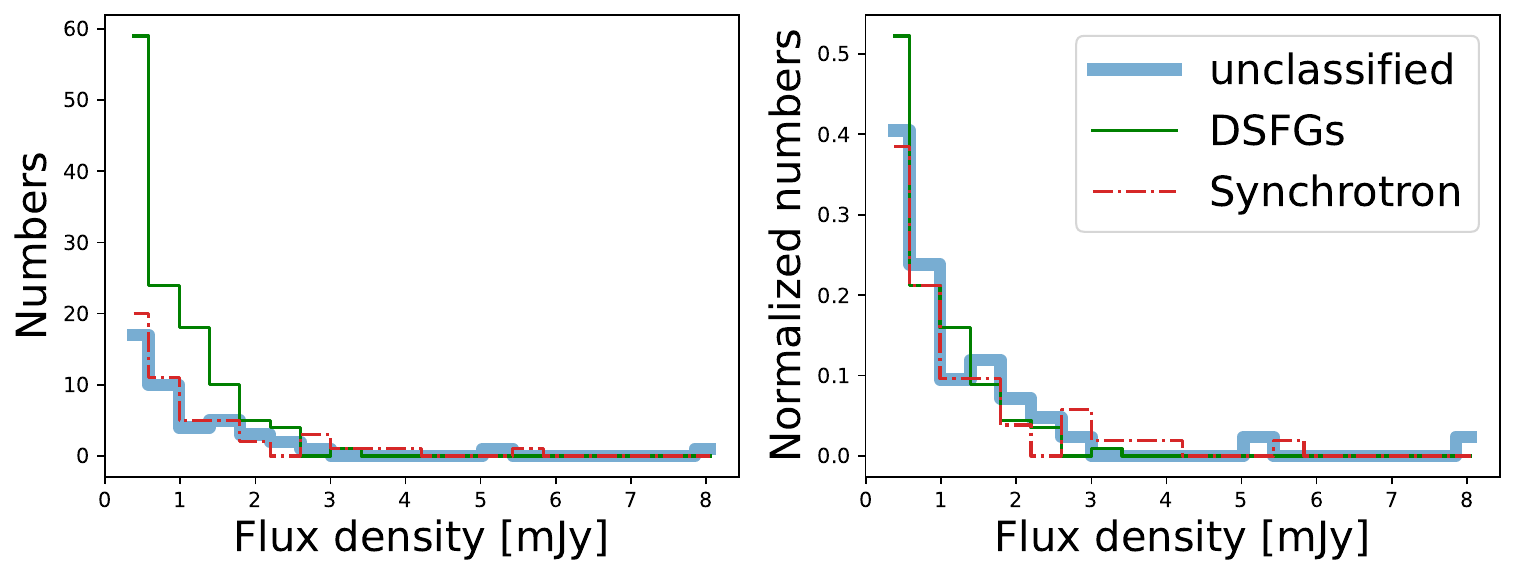}
   \includegraphics[width=\linewidth]{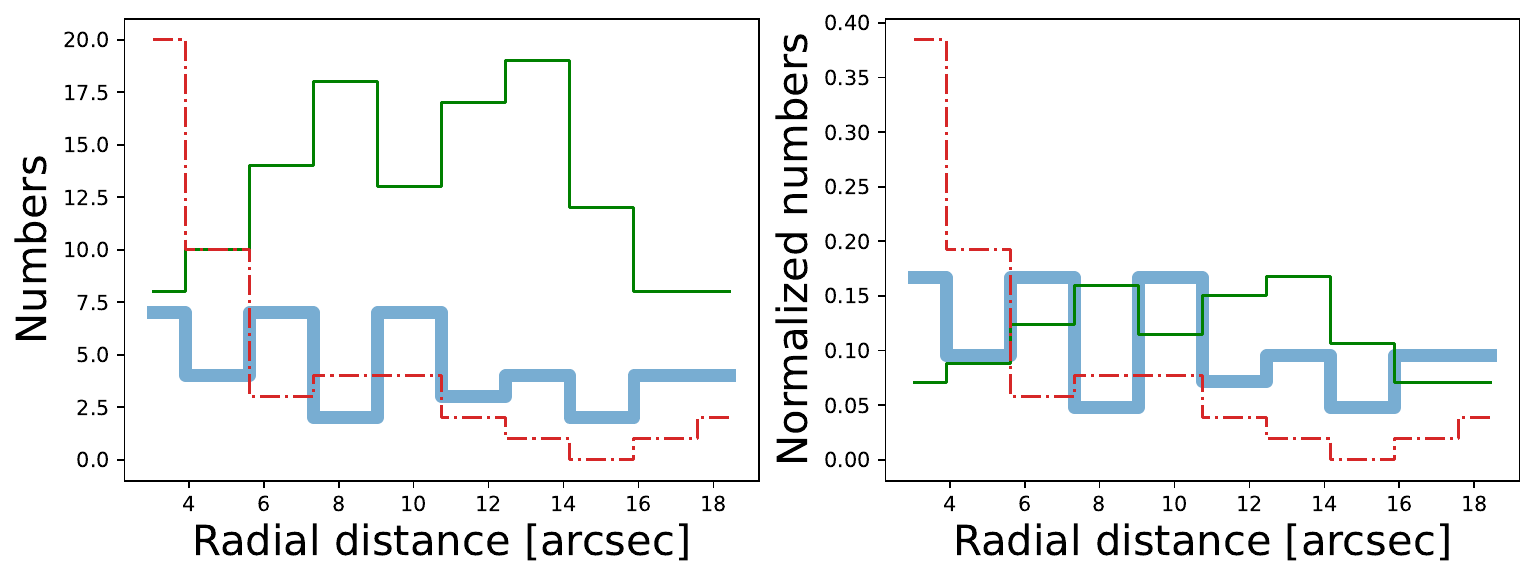}
   \caption{Histograms of all the detections in flux bins and radial distance bins from band 6. The first row shows the number and the normalized number of detections that change with the flux density. We do not find an evident difference in the flux density distribution of the three classifications. The second row shows the number and the normalized number of detections change with their radial distance to the central calibrator. Synchrotron sources concentrate near the calibrator, which is consistent with the nature of radio jets associated with the blazar. Compared with Synchrotron, DSFGs have a flatter distribution as a function of radius. Due to the small number of unclassified sources, we can not classify them by their radial distance, but including them and excluding them will not change our main result.}%
   \label{appendixfig:B6_histogram}
\end{figure}

The radial distance and flux distribution of all the detections can offer additional information about the accuracy of the source classification.
In Fig.~\ref{appendixfig:B6_histogram}, we show the histograms of band 6 sources, where the number of detections is largest, in different bins of flux and radial distance.
All three categories share a similar trend in different flux bins, but they are different in the distribution of radial distance.
Most of the radio sources are quite close to the calibrator ($<$5\,arcsec), which should be the natural outcome if the majority of them are blobs of radio jets.
Similar to our prediction in \ref{sec:selection_bias}, we have more DSFGs at the median radial distance.
The unclassified sources have a much flatter distribution, which should include both RSs and DSFGs.
However, their relatively small number prohibits a quantitive analysis.
In the next section, we will randomly sample these unclassified sources to test their influence on the final number counts.

\begin{figure}
   \centering
   \includegraphics[width=0.9\linewidth]{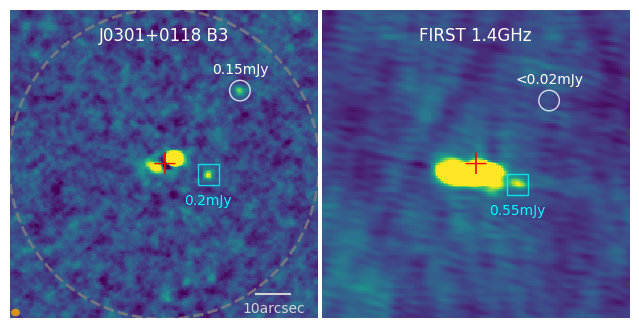}
   \includegraphics[width=0.9\linewidth]{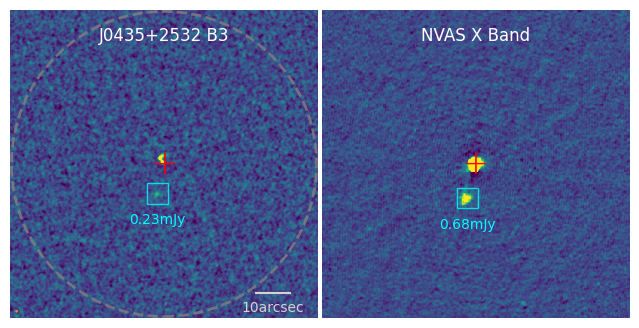}
   \includegraphics[width=0.9\linewidth]{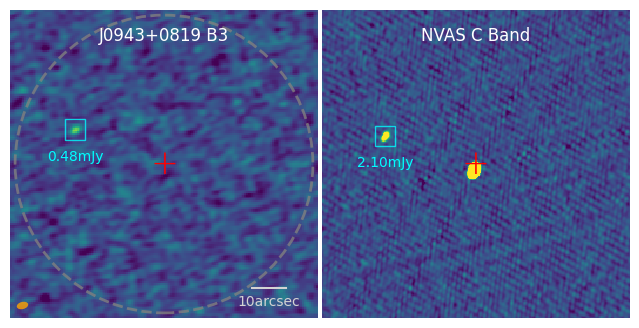}
   \caption{Examples of confirmed radio sources from VLA archive images. The first column shows the band 3 images from ALMACAL. The second column shows the corresponding radio images from VLA archive and they have been cropped to the same scale as the ALMACAL images. The dashed grey circle is the FoV adopted in this work, which is 1.8$\times$FWHM of the respective primary beam}. SMG is marked with white circle and synchrotron emission is marked with cyan square. Radio images are useful to constrain the emission mechanisms of ALMACAL detections without multi-bands observations.%
   \label{appendixfig:VLA_and_ALMA}
\end{figure}

\section{Number counts in the inner and outer parts of the fields}\label{appdixsec:number_counts_innerouter}
\begin{figure}
   \centering
   \includegraphics[width=0.9\linewidth]{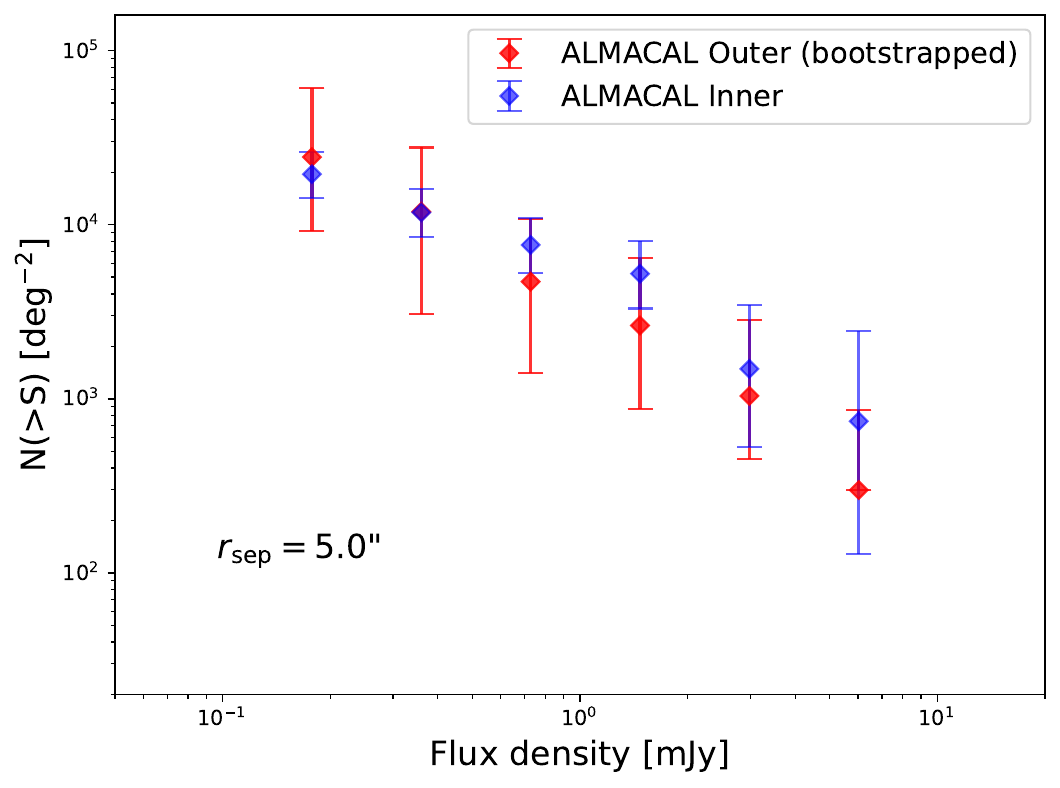}
   \includegraphics[width=0.9\linewidth]{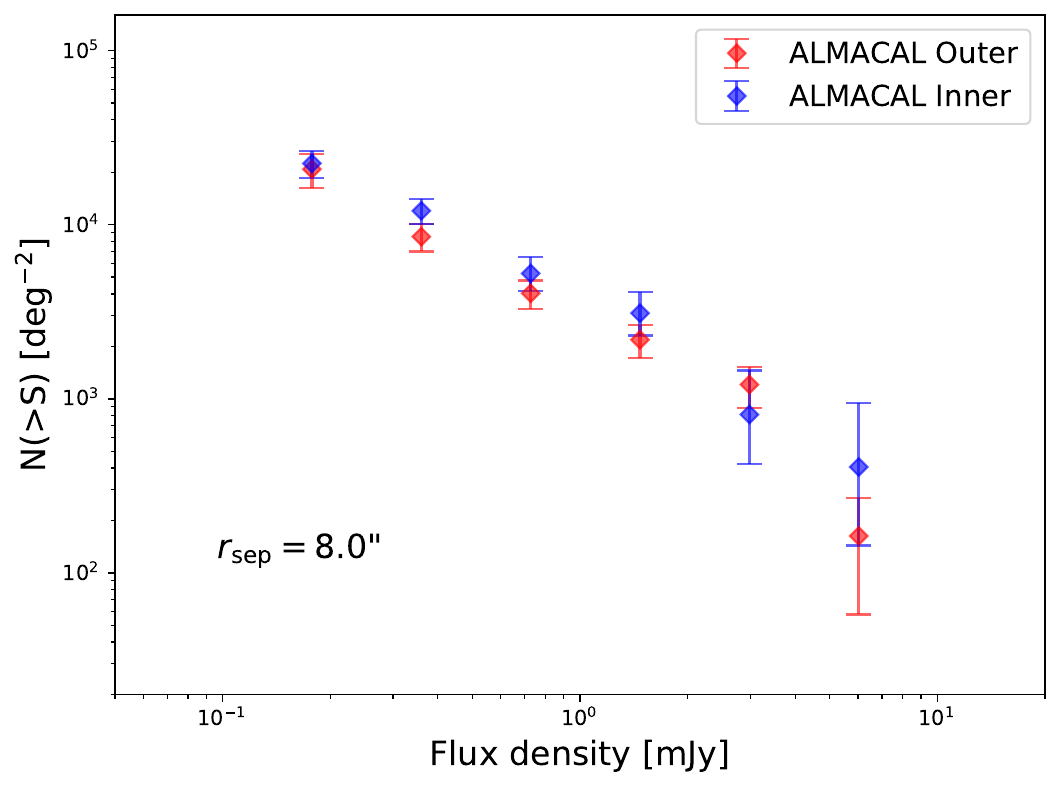}
   \caption{The comparison between the number counts in the inner parts and the outer parts of all the fields. We use all the inner regions but randomly sample the available outer regions to reach the same total sky coverage as the inner ones. The difference in number counts between the inner and outer regions is well within their intrinsic uncertainties.}%
   \label{appendixfig:inner_outer}
\end{figure}

The dark matter halo associated with the central calibrator can also act as a gravitational lens.
It can stretch the background DSFGs and produce multiple images, which can potentially boost the number counts in the inner regions.
We test this effect in band 7, where the effect is the strongest.
We first divide our fields into the inner and outer regions defined by the separation radius $r_{\rm sep}$. 
We then calculate the number count in the inner and outer regions separately.
If the total area of the inner regions is significantly smaller than the outer regions, we bootstrap the outer regions that sample the same total area as the inner regions.
We show the difference between inner and outer regions in Fig.~\ref{appendixfig:inner_outer}.
We see no statistically significant evidence for excess source counts in the inner 5--8$''$ regions, as might be caused by either lensing from a halo associated with the calibrator, or from clustering or interactions of the SMGs around the blazar.

\section{Robustness of number counts}\label{appdixsec:number_counts_robustness}
\begin{figure*}
  \centering
  \includegraphics[width=.45\linewidth]{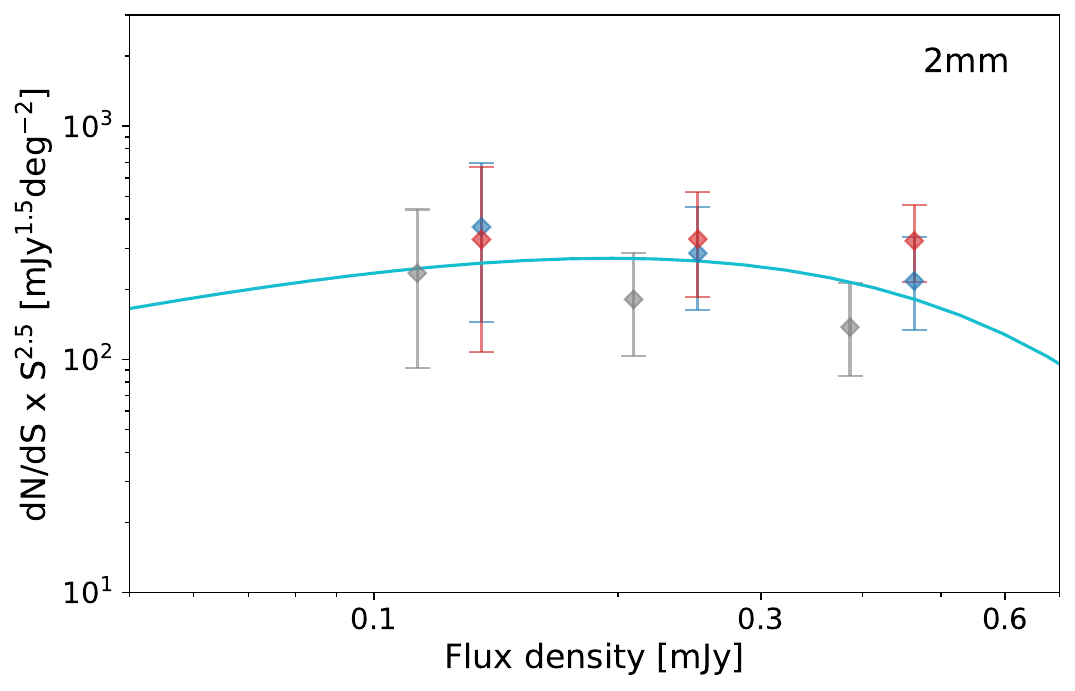}
  \includegraphics[width=.45\linewidth]{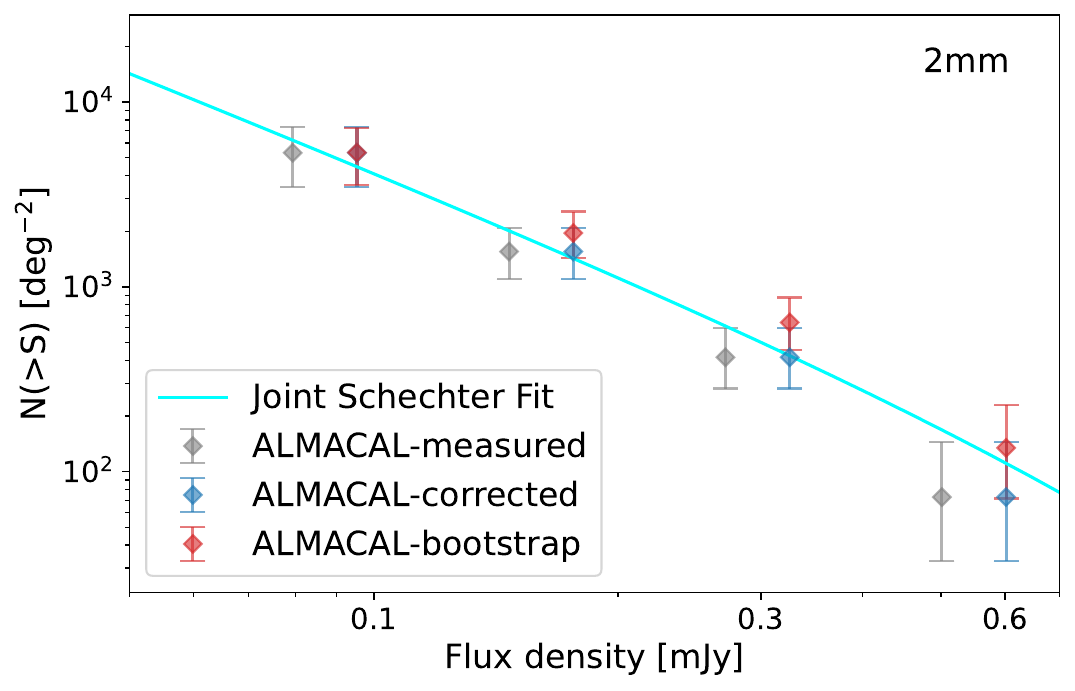}
  \includegraphics[width=.45\linewidth]{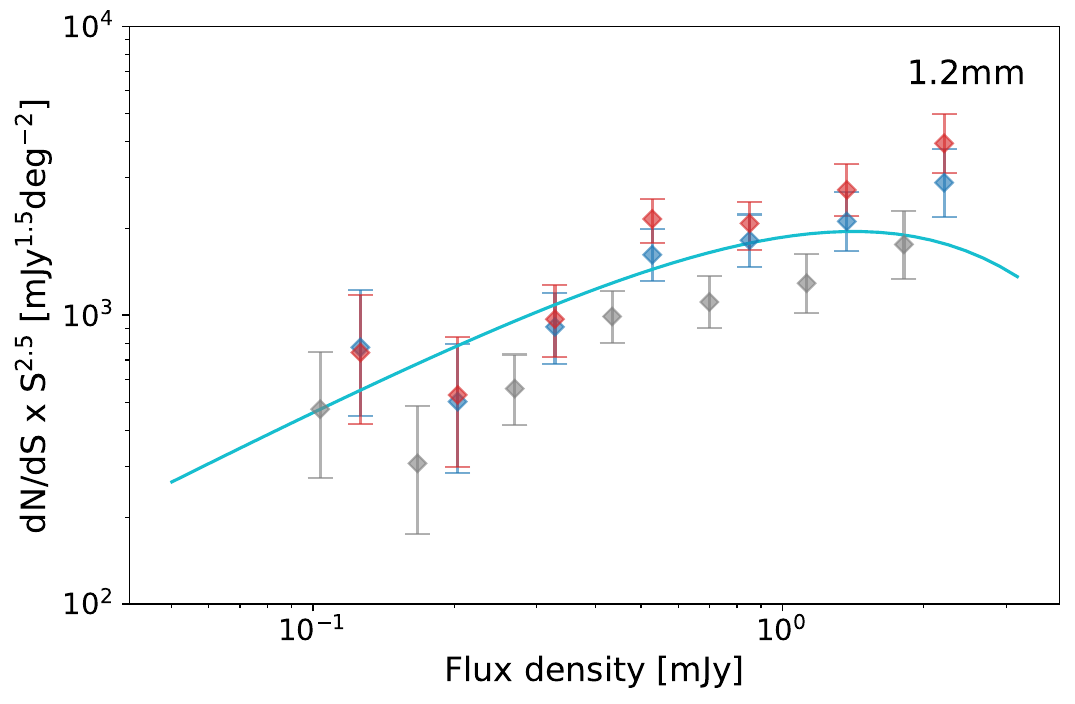}
  \includegraphics[width=.45\linewidth]{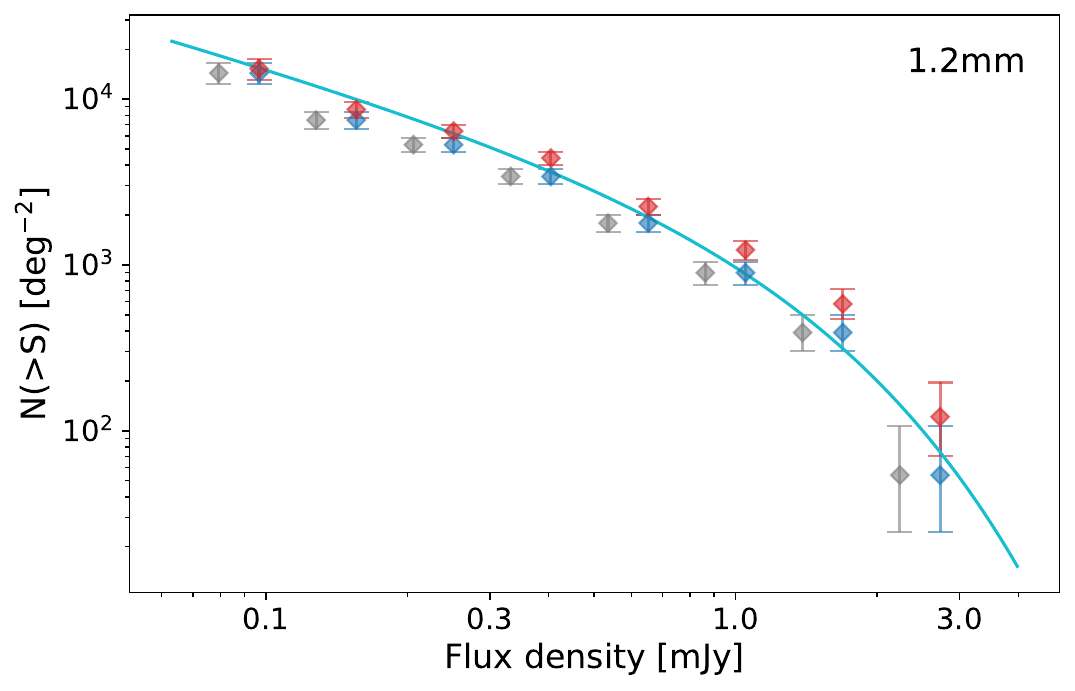}
  \includegraphics[width=.45\linewidth]{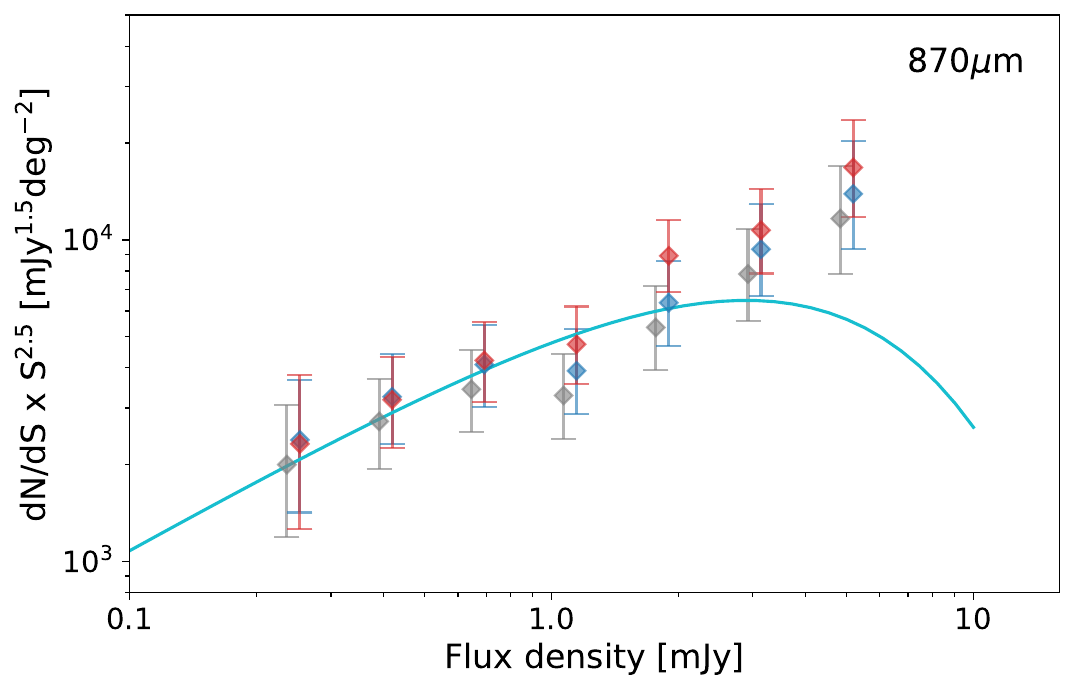}
  \includegraphics[width=.45\linewidth]{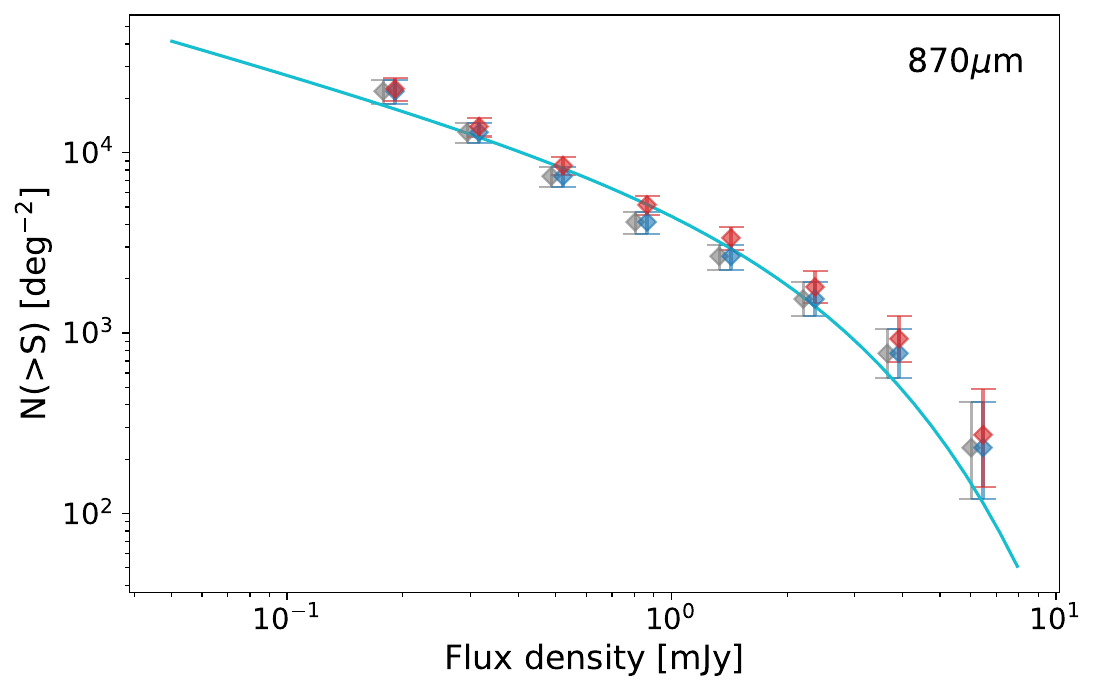}
  \includegraphics[width=.45\linewidth]{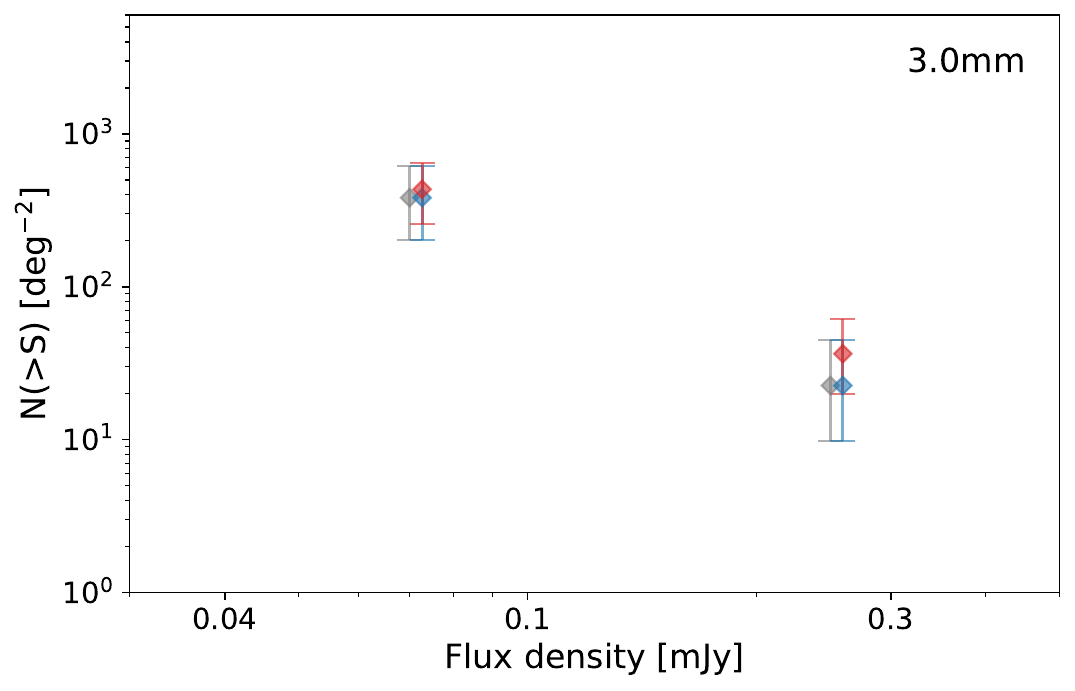}
  \includegraphics[width=.45\linewidth]{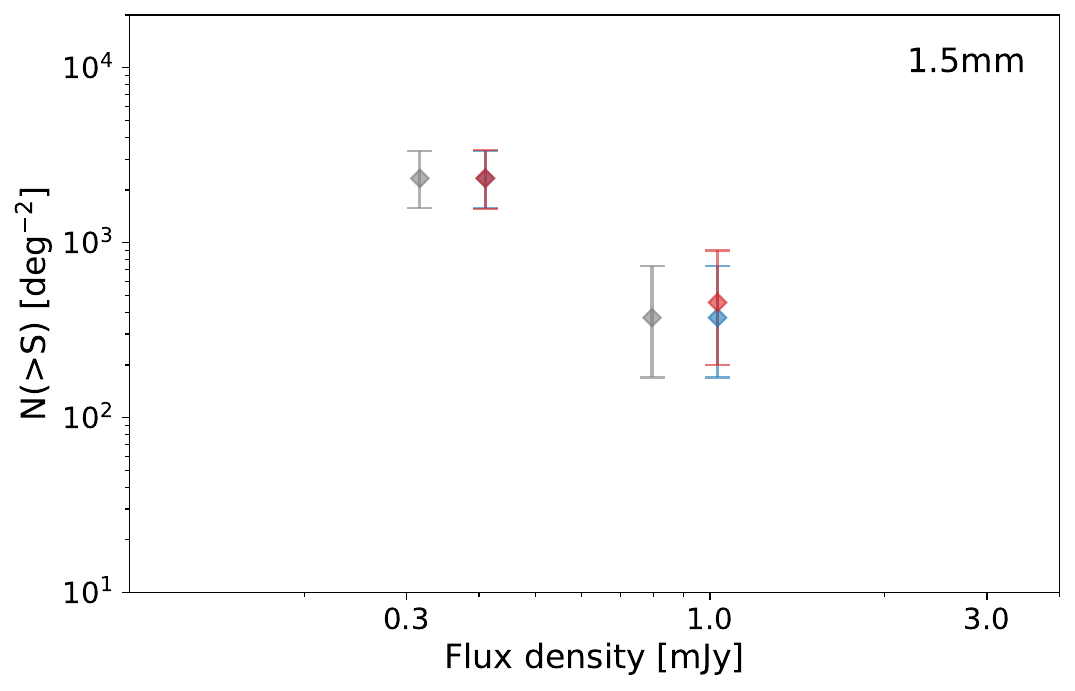}
  \caption{The stability of number counts. We show the final results reported in \S\ref{sec:number_counts} along with the results before the effective wavelength correction (see more in \S\ref{sec:effective_wavelength}) and the bootstrapped number counts in each band (see more in \S\ref{appdixsec:number_counts_robustness}). Bootstrapping is achieved by randomly sampling the unclassified sources into the final DSFG sample based on their radial distribution. The bootstrapped number counts are consistent with the original results within the uncertainties, which indicates our number counts are robust against unclassified sources. We also show the best-fitt Schechter functions reported from Fig.~\ref{fig:number_counts_B7} to Fig.~\ref{fig:number_counts_B4}.}
  \label{appendixfig:number_counts_perturbation}
\end{figure*}

To quantify the contribution of these unclassified detections to our final number counts, we calculate their probability of being radio sources or DSFGs based on the overall number of confirmed sources in each band.
As we see in \S\ref{appendixfig:B6_histogram}, DSFGs have different radial distribution as radio sources, we thus calculate the probability of being a DSFG at different radial bins.
For instance, the probability that a single-band detection in band 6 is a DSFG is equal to the fraction of DSFGs amongst all the classified sources in band 6 at the same radial distance.
We choose the radial bins to have at least 10 sources in each bin.
We repeat this process $1,000$\,times to capture the variation of the final number counts.
The comparison of the bootstrapped number counts and the original number counts is in Fig.~\ref{appendixfig:number_counts_perturbation}.
The bootstrapped number counts are slightly larger than the number counts of the confirmed DSFGs sample, but they are still consistent with the original number counts within the uncertainties.

\end{document}